\pdfoutput=1
\documentclass[nonacm, screen]{acmart}
\setcopyright{acmcopyright}

\usepackage{microtype}
\usepackage{array}
\usepackage{booktabs}
\usepackage{pgffor}
\usepackage{caption}
\usepackage{hyperref}

\usepackage{dblfloatfix}

\usepackage{xspace}
\usepackage{arcs}
\usepackage{multirow}

\newcommand{\replicability}{\texttt{REPLICABILITY.md}\xspace}
\newcommand{\readme}{\texttt{README.md}\xspace}

\newlength\Colsep
\newcommand{\vc}{\textsc{VC}\xspace}
\newcommand{\vcfull}{\textsc{Vertex Cover}\xspace}
\newcommand{\myoct}{\textsc{OCT}\xspace}
\newcommand{\myoctfull}{\textsc{Odd Cycle Transversal}\xspace}

\newcommand{\ilp}{\textsc{ILP}\xspace}
\newcommand{\ilpfull}{\textsc{Integer Linear Programming}\xspace}
\newcommand{\qubo}{\textsc{QUBO}\xspace}
\newcommand{\qubofull}{\textsc{Quadratic Unconstrained Binary Optimization}\xspace}
\newcommand{\huffner}{H{\"u}ffner\xspace}

\newcommand{\wh}{\texttt{WH}\xspace}
\newcommand{\beasley}{\texttt{Beasley}\xspace}
\newcommand{\gka}{\texttt{GKA}\xspace}
\newcommand{\bfs}{\textsf{BFS}\xspace}
\newcommand{\dfs}{\textsf{DFS}\xspace}
\newcommand{\luby}{\textsf{Luby}\xspace}
\newcommand{\mindeg}{\textsf{MinDeg}\xspace}
\newcommand{\netx}{\textsf{NetworkX}\xspace}
\newcommand{\er}{Erd\"os-R\'enyi\xspace}
\newcommand{\tunoct}{Tunable-OCT\xspace}
\newcommand{\ba}{Barab\'asi-Albert\xspace}
\newcommand{\cl}{Chung-Lu\xspace}
\newcommand{\cplex}{\textsf{CPLEX}\xspace}
\newcommand{\glpk}{\textsf{GLPK}\xspace}

\usepackage[T1]{fontenc}
\newsavebox\CBox
\def\textBF#1{\sbox\CBox{#1}\resizebox{\wd\CBox}{\ht\CBox}{\emph{\color{blue} #1}}}

\newenvironment{tightcenter}
 {\parskip=0pt\par\nopagebreak\centering}
 {\par\noindent\ignorespacesafterend}

\usepackage{tikz}
\usetikzlibrary{calc}

\usepackage{xspace}
\usepackage{xargs}
\usepackage{xifthen}
\usepackage{amsfonts}
\usepackage{amsmath}
\usepackage{framed}

\usepackage{ctable}
\newlength{\RoundedBoxWidth}
\newsavebox{\GrayRoundedBox}
\newenvironment{GrayBox}[1]%
{\setlength{\RoundedBoxWidth}{\linewidth-4.5ex}
\def\boxheading{#1}
\begin{lrbox}{\GrayRoundedBox}
\begin{minipage}{\RoundedBoxWidth}%
}{%
\end{minipage}
\end{lrbox}%
\begin{tightcenter}%
\begin{tikzpicture}%
\node(Text)[draw=black!20,fill=white,rounded corners,%
inner sep=2ex,text width=\RoundedBoxWidth]%
{\usebox{\GrayRoundedBox}};
\coordinate(x) at (current bounding box.north west);
\node [draw=white,rectangle,inner sep=3pt,anchor=north west,fill=white]
at ($(x)+(10.5pt,.75em)$) {\boxheading};
\end{tikzpicture}
\end{tightcenter}\vspace{0pt}%
\ignorespacesafterend
}

\newenvironment{problem}[2][]{\noindent\ignorespaces%
\FrameSep=6pt%
\parindent=0pt%
\vspace*{-.5em}
\ifthenelse{\isempty{#1}}{%
\begin{GrayBox}{\textsc{#2}}%
}{%
\begin{GrayBox}{\textsc{#2} parametrised by~{#1}}%
}
\newcommand\Prob{Problem:}%
\newcommand\Input{Input:}%
\newcommand\Objective{Objective:}%
\begin{tabular*}{\columnwidth}{@{\hspace{.5em}} >{\itshape} p{1.4cm} p{0.85\columnwidth} @{}}%
}{
\end{tabular*}%
\end{GrayBox}%
\vspace*{-.5em}
\ignorespacesafterend
}

\newenvironment{probbox}[1]{\noindent\ignorespaces%
\FrameSep=6pt%
\parindent=0pt%
\vspace*{-.5em}
\begin{GrayBox}{\textsc{#1}}%
}{
\end{GrayBox}%
\vspace*{-.5em}
\ignorespacesafterend
}

\begin{document}

\title{An Updated Experimental Evaluation of Graph Bipartization Methods}

\author{Timothy D. Goodrich}
\affiliation{
\institution{North Carolina State University}
\streetaddress{890 Oval Dr}
\city{Raleigh}
\state{NC}
\postcode{27606}
\country{USA}
}
\email{tdgoodri@ncsu.edu}

\author{Eric Horton}
\affiliation{
\institution{North Carolina State University}
\streetaddress{890 Oval Dr}
\city{Raleigh}
\state{NC}
\postcode{27606}
\country{USA}
}
\email{ewhorton@ncsu.edu}

\author{Blair D. Sullivan}
\affiliation{
\institution{North Carolina State University}
\streetaddress{890 Oval Dr}
\city{Raleigh}
\state{NC}
\postcode{27606}
\country{USA}
}
\email{blair\_sullivan@ncsu.edu}
\affiliation{
\institution{University of Utah}
\streetaddress{201 Presidents' Cir}
\city{Salt Lake City}
\state{UT}
\postcode{84112}
\country{USA}
}
\email{sullivan@cs.utah.edu}

\renewcommand\shortauthors{Goodrich, T.D. et al}

\begin{abstract}
We experimentally evaluate the practical state-of-the-art in graph bipartization (\myoctfull), motivated by recent advances in near-term quantum computing hardware and the related embedding problems.
We assemble a preprocessing suite of fast input reduction routines from the \myoctfull (\myoct) and \vcfull (\vc) literature, and compare algorithm implementations using \qubofull problems from the quantum literature. We also generate a corpus of frustrated cluster loop graphs, which have previously been used to benchmark quantum annealing hardware. The diversity of these graphs leads to harder \myoct instances than in existing benchmarks.

In addition to combinatorial branching algorithms for solving OCT directly, we study various reformulations into other NP-hard problems such as \vc and \ilpfull (\ilp), enabling the use of solvers such as \cplex.
We find that for heuristic solutions with time constraints under a second, iterative compression routines jump-started with a heuristic solution perform best, after which point using a highly tuned solver like \cplex is worthwhile.
Results on exact solvers are split between using \ilp formulations on \cplex and solving \vc formulations with a branch-and-reduce solver.
We extend our results with a large corpus of synthetic graphs, establishing robustness and potential to generalize to other domain data.
In total, over 8000 graph instances are evaluated, compared to the previous canonical corpus of 100 graphs.

Finally, we provide all code and data in an open source suite, including a Python API for accessing reduction routines and branching algorithms, along with scripts for fully replicating our results.

\end{abstract}

%
%


%
%


\maketitle

\clearpage
\section{Introduction}
\label{section:introduction}
\myoctfull (\myoct), the problem of deleting vertices to make a graph bipartite, has been well-studied in the theory community over the last two decades.
Techniques such as \emph{iterative compression} \cite{reed2004finding, huffner2009algorithm} and \emph{branch-and-reduce} \cite{lokshtanov2014faster, akiba2016branch} have led to significant improvements in both worst-case and experimental run times.
These improvements are most drastically seen on the canonical \myoct benchmark, Wernicke's \textsc{Minimum Site Removal} dataset \cite{wernicke2003algorithmic} (denoted \wh), where run times have dropped from over 10 hours \cite{wernicke2003algorithmic} to under 3 minutes \cite{huffner2009algorithm} to under 1 second per instance for multiple state-of-the-art solvers \cite{akiba2016branch}.
While these results illustrate the rapid algorithmic advances, they also show that new data is needed for further study.

Recently, a need for practical graph bipartization algorithms has arisen in quantum computing, where the hardware and/or problem structure may naturally have underlying bipartite structure that can be exploited algorithmically\footnote{For example, the D-Wave Chimera hardware is bipartite, and upcoming Pegasus hardware admits a large complete bipartite graph embedding.}. In contrast to the \wh data -- which was expected to be bipartite barring read errors -- problem instances from quantum annealing may not be close to bipartite. Additionally, in the quantum setting ``good enough'' solutions are of interest since \myoct may be solved as a subroutine in a larger automated compiler. This introduces a run time vs. solution quality trade-off previously not considered. In this work, we combine richer data with various timeout scenarios in order to provide a modernized evaluation of \myoct methods.

\subsection{Related Work}
\label{subsection:related_work}
Modern theoretical advances on \myoct began with the seminal result of Reed, Smith, and Vetta \cite{reed2004finding}, who showed that the problem -- which asks one to remove the minimum number of vertices $k$ to produce a bipartite subgraph -- is fixed-parameter tractable with parameter $k$ using the technique of \emph{iterative compression}. This algorithm was initially shown to run in time $O(4^k km)$, but improved analyses showed a $O(3^k km)$ run time and simpler algorithms for the compression routines~\cite{huffner2009algorithm, lokshtanov2009simpler}.
The next theoretical improvement came from an improved algorithm and analysis for \vcfull (\vc) and a (straightforward) conversion of an \myoct instance to a \vc instance.
This strategy results in an $O^*(2.3146^{k'})$ algorithm,\footnote{$O^*(f(k))$ denotes $O(f(k) n^c)$ for some constant $c$} where $k'$ denotes the gap between an optimal solution to \vcfull and the solution given by the linear programming (LP) relaxation \cite{lokshtanov2014faster}.
Recent work has used the half-integrality of LP-relaxations to reduce the polynomial in $n$ at the cost of a higher parameterized term, resulting in an $O(4^k n)$ algorithm for \myoct \cite{iwata2014linear}, and $O^*(4^k)$ \cite{wahlstrom2017lp} and $O(4^k n)$ \cite{iwata20170} algorithms for the more general problem of \textsc{Non-Monotonic Cycle Transversal}.
Other algorithmic results for \myoct include a $O(\log{\sqrt{n}})$-approximation algorithm \cite{agarwal2005log}, a randomized algorithm based on matroids \cite{kratsch2014compression}, and a subexponential algorithm on planar graphs \cite{lokshtanov2012subexponential}.

On the practical side, the first implementation was a branch-and-bound algorithm by Wernicke in 2003 \cite{wernicke2003algorithmic} used for studying single nucleotide polymorphisms.
A greedy depth-first search heuristic was used to identify upper bounds on \myoct, and several sparse graph reduction routines were applied before branching.
A Java implementation of this algorithm solved most \wh instances within a 10 hour timeout.
In 2009, \huffner implemented a refined iterative compression algorithm~\cite{huffner2009algorithm} with additional pruning for infeasible assignments and symmetry in the compression routine's branching algorithm to achieve experimentally faster run times; all of the \wh instances could then be solved within three minutes.
\huffner compared this algorithm against an \ilp formulation using the GNU Linear Programming Kit (\textsf{GLPK}) \cite{glpk}, which had unfavorable run times.
More recently, Akiba and Iwata \cite{akiba2016branch} used a \vc-solver based on branch-and-reduce to solve \myoct using a standard transformation to \vc \cite{lokshtanov2014faster}.
The authors reported that their open source Java implementation could solve all \wh data within a second, while competing implementations based on maximum clique and an \ilp formulation solved using \textsf{CPLEX} \cite{cplex} all finished within three seconds.

\subsection{Our Contributions}
In this work, we collect existing \myoct techniques, provide a common Python API for running these algorithms, and evaluate them within a broader experimental envelope, incorporating quantum-inspired data and use cases. We also provide a frame for generalizing our conclusions with a large synthetic corpus representing a variety of random graph models.

Whereas \myoct can be computed within seconds for all graphs in the previous \wh dataset~\cite{wernicke2003algorithmic, huffner2009algorithm}, we provide a new, significantly more difficult benchmark corpus.
Motivated by the widespread usage of \qubofull (\qubo) problems in quantum annealing research~\cite{neven2008image}, we select \qubo instances from a recent survey~\cite{dunning2015works} that would be of interest to practitioners working on near-term quantum annealers.
These datasets are selections from Glover, Kochenberger, and Alidaee \cite{glover1998adaptive} (denoted \gka) and Beasley \cite{beasley1998heuristic} (denoted \beasley).
Not only do these datasets contain significantly harder \myoct instances, they also represent a wider array of graph properties such as number of vertices, edge density, degree distribution, etc., than the \wh benchmark.

Collecting previous code and providing missing implementations, we assemble a unified Python API allowing easy comparison of prior work.
Preprocessing routines are taken from the \myoct \cite{wernicke2003algorithmic} and \vc literature \cite{akiba2016branch} and applied to all datasets to harden the benchmark corpus.
Heuristics for \myoct upper bounds \cite{wernicke2003algorithmic, goodrich2017optimizing} are collected and implemented as a standalone heuristic ensemble solver for stochastically sampling `good enough' solutions.
We use these heuristic solutions, along with a density-first heuristic for the compression ordering, to jump-start the iterative compression algorithm of \huffner \cite{huffner2009algorithm}.
These combinatorial algorithms for solving \myoct directly are complemented by \vc-based \cite{akiba2016branch} and \ilp-based \cite{cplex} solvers.

In quantum-specific experiments, we examine two distinct use cases.
First, to represent scenarios where an automated compiler may use \myoct as a subroutine and accept heuristic solutions, we evaluate the heuristic ensemble, iterative compression, and the \ilp formulation under timeouts of 0.01, 0.1, 1, and 10 seconds.
We find that the iterative compression implementation jump-started with heuristic solutions performs best for timeouts less than a second, after which it is worth paying the overhead of using an \ilp solver such as CPLEX.
In a second use case where an exact solution is required in order to recognize un-embeddable quantum programs, we evaluate iterative compression, \vc-based, and \ilp-based exact solvers.
Here we again find that \ilp formulations solved by CPLEX dominate, typically by at least an order of magnitude.

Generalizing these results, we generate synthetic graphs using four random graph generators -- including the frustrated cluster loop model, which has been previously used to generate instances for quantum annealing -- and
evaluate whether ``generic'' instances matching the density, degree distribution, etc. exhibit
similar effectiveness of reduction routines and solver run times.
We find that our results on \qubo data are robust, with the same best practice recommendations.
This experimental evidence provides practitioners with a useful reference for developing custom solutions for particular applications.

Our work is fully replicable, with documented code ~\cite{practical-oct} open sourced under the BSD 3-Clause license.
For the interested reader, Appendix A contains implementation details and Appendix B contains extended results.

\section{Background}
\label{section:background}

We denote a graph $G = (V, E)$. For a set of vertices $S$, we denote the subgraph induced by deleting $S$ as $G \setminus S$. An edge $(u, v) \in E$ can be \emph{contracted} by adding a new node $uv$, adding edges from $uv$ to all neighbors of $u$ and $v$, then deleting $u$ and $v$. A graph $P$ is a \emph{minor} of a graph $H$ if $P$ can be obtained from $H$ with vertex deletion, edge deletion, and edge contraction. In the context of quantum annealing, we denote the graph $P$ as a \emph{problem graph} and $H$ as a \emph{hardware graph}.

\myoctfull (\myoct) is formally defined as an optimization problem with natural parameter $k$.

\begin{minipage}{0.6\textwidth}
\vspace{0.1in}
\begin{problem}{\myoctfull (\myoct)}
\Input & An input graph $G=(V,E)$.\\
\Prob  & Find $S \subseteq V$ such that $G \setminus S$ is bipartite.\\
\Objective & Minimize $k := |S|$.
\end{problem}
\end{minipage}

\myoct is closely related to \vcfull (\vc).

\begin{minipage}{0.6\textwidth}
\vspace{0.1in}
\begin{problem}{\vcfull (\vc)}
\Input & An input graph $G=(V,E)$.\\
\Prob  & Find $S \subseteq V$ such that $G \setminus S$ is edgeless.\\
\Objective & Minimize $|S|$.
\end{problem}
\end{minipage}

Specifically, given a graph $G$, we create an auxiliary graph $G'$ with two copies of $G$ (which we call left and right) joined by a matching between corresponding vertices. Formally, $G' = (V_L \cup V_R, E_L \cup E_R \cup E')$, where for $i \in \{L, R\}$, $V_i = \{v_i ~|~ v \in V\}$ and $E_i = \{(u_i, v_i) ~|~ (u, v) \in E\}$, and $E' = \{(v_L, v_R) ~|~ v \in V\}$ is the matching edges.
A solution $S'$ for \vc on $G'$ can be mapped to a solution $S$ for \myoct on $G$ with $S = \{v ~|~ v_L \in S \text{ and } v_R \in S\}$; this mapping preserves optimality \cite{akiba2016branch}.

Both \myoct and \vc can also be rewritten as \ilpfull (\ilp) instances; as explored further in Section~\ref{section:alternative_solvers}, the choice of $\myoct \to \ilp$ formulation has performance implications.

\begin{minipage}{0.6\textwidth}
\vspace{0.1in}
\begin{minipage}{\linewidth}
\begin{probbox}{\myoctfull (\ilp) \cite{huffner2009algorithm}}
\emph{Input:} \hskip1em $G=(V,E)$.
\begin{align*}
\text{Minimize} & & \sum_{v \in V} c_v & & \\
\text{s.t.} & & s_v + s_u + c_v + c_u \geq 1 & & \forall (u, v) \in E\\
            & & s_v + s_u -  c_v - c_u \leq 1 & & \forall (u, v) \in E\\
            & & s_v \in \{0, 1\} & & \forall v \in V\\
            & & c_v \in \{0, 1\} & & \forall v \in V
\end{align*}
\end{probbox}
\end{minipage}
\end{minipage}

\begin{minipage}{0.6\textwidth}
\vspace{0.1in}
\begin{minipage}{\linewidth}
\begin{probbox}{\vcfull (\ilp) \cite{akiba2016branch}}
\emph{Input:} \hskip1em $G=(V,E)$.
\begin{align*}
\text{Minimize} & & \sum_{v \in V} x_v & & \\
\text{s.t.} & & x_u + x_v \geq 1 & & \forall (u, v) \in E\\
            & & x_v \in \{0, 1\}& & \forall v \in V
\end{align*}
\end{probbox}
\end{minipage}
\end{minipage}

An \emph{anytime algorithm} will return the best solution found so far when given a timeout signal.
An \emph{exact algorithm} will return a (provably) optimal solution, whereas a \emph{heuristic} is unable to prove optimality.
A middle ground between heuristic and exact is an \emph{approximation algorithm}, whose error is bound with a constant.
An $\alpha$-approximation algorithm for a minimization problem has solution size $k$ bounded by $k \leq \alpha \cdot OPT$, where $OPT$ is the optimal solution size; $\alpha$ is the \emph{approximation factor}.
We define the approximation factor over a corpus of graph instances as the worst factor over all individual instances in the corpus.

\subsection{Graph bipartization in quantum annealing}
Recent advances in quantum computing has resulted in a class of optimization problem solvers denoted \emph{quantum annealers} (QA).
Implementations of QA hardware are based off of the adiabatic quantum annealing (AQC) model, but are typically less powerful (i.e. not universal quantum computers).
These QA devices have \emph{hardware topologies}, these graphs define the communication pathways (edges) between qubits (vertices).
For example, D-Wave Systems currently produces QA hardware with Chimera hardware, and is preparing for a second generation Pegasus model.

Graph bipartization (\myoct) occurs during the compilation step in quantum annealing, when a problem graph representing the optimization problem being solved by the quantum computer\footnote{e.g. a \qubofull (\qubo)} must be \emph{embedded} as a minor into the graph representing the quantum hardware~\cite{choi2011minor}.
Researchers have had success running naturally bipartite QUBOs (e.g. deep learning models) on D-Wave Systems annealers \cite{schrock2017recall, kosko1988bidirectional,boyda2017deploying, potok2018study}.
Generalizing these tools to non-bipartite QUBOs is currently of interest to enable additional applications (e.g. Karp's 21 NP-hard problems \cite{lucas2014ising}).
The area of automatic embedding tools is under active development (refer to, e.g., \cite{goodrich2017optimizing, hamilton2017identifying, venturelli2015quantum}).

We examine two distinct use cases of graph bipartization in quantum program compilation.

First, a feasible bipartization of a graph is useful for certain embedding algorithms \cite{goodrich2017optimizing, hamilton2017identifying}, providing a structure on which to limit the search space of heuristics.
In this use case, an automated compiler might require this bipartization in as little as 0.01 seconds, whereas a computer-assisted researcher working in an interactive environment may wait closer to 10 seconds.

Second, \myoct can be used to identify when a particular program cannot embed into hardware, as is shown for D-Wave Systems' \emph{Chimera} hardware in \cite{goodrich2017optimizing}.
This scenario requires that the solver return a certified optimal \myoct solution, but longer run times are permissible since a hardware owner can compute forbidden configurations once per hardware model.
We examine both of these use cases in more detail in Section \ref{section:quantum_results}.

\section{Algorithm Overview}
\label{section:algorithm_overview}
In this section, we overview various algorithmic techniques previously applied to \myoct.
We begin with reduction routines from both the \myoct and \vc literature, then continue to linear-time heuristics historically used to provide upper bounds for branch-and-bound approaches.
The first exact solver we detail is \huffner's iterative compression solver~\cite{huffner-ic}, which we enhance with heuristics to create an anytime algorithm with increased performance particularly when given small timeouts.
Finally, we detail how CPLEX can solve \myoct using various reformulations into \ilpfull.

\subsection{Reduction Routines}
\label{section:reduction_routines}
We begin with \emph{reduction routines} -- rules for simplifying the graph instance such that a solution on the reduced instance is valid for the original instance.
Typically these reductions will remove vertices by recognizing configurations that can be simplified deterministically.
Reduction routines for \myoct come from two sources.
Wernicke's branch-and-bound algorithm \cite{wernicke2003algorithmic} uses nine reductions directly on the \myoct instance.
These reductions form roughly three categories: removal of standalone structures (e.g., bipartite components and degree-1 vertices), removal of vertex separators which induce certain bipartite components, and reconfigurations of local structures (e.g., removing a degree-2 vertex in an induced four-cycle).
These reductions are most effective on close-to-bipartite graphs, low-connectivity graphs, and sparse graphs, respectively.

A second source of reductions is Akiba and Iwata's \vc solver \cite{akiba2016branch}, which uses nine reduction routines specific to the \vc instance's graph.
Based on the conversion between \myoct and \vc in Section~\ref{section:background}, a vertex $v$ in the \myoct instance must be in the transversal if both $v_1$ and its mirror $v_2$ must be in the vertex cover in the \vc instance.
Similarly, $v$ is labeled bipartite if $v_1$ or $v_2$ is excluded from an optimal vertex cover.

We refer the interested reader to the original papers \cite{wernicke2003algorithmic, akiba2016branch} for detailed definitions, examples, and complexity analysis.
We provide an open source implementation of Wernicke's reductions using Python and NetworkX, and modify a copy of Akiba and Iwata's \vc solver \cite{ai-code} to output the graph after a single round of reductions.

\subsection{Heuristics and Approximations}
\label{subsection:heuristics}
Heuristics for \myoct typically compute a maximal bipartite induced subgraph, then label
all remaining vertices as an odd cycle transversal.
One strategy for finding a large bipartite subgraph is greedily 2-coloring the vertices
using a depth-first search, and adding incompatible vertices to an OCT set as needed~\cite{wernicke2003algorithmic}; this heuristic has a natural breadth-first search variant.
Both of these methods are nondeterministic with respect to the choice of the initial vertex and the order in which neighbors are added to the search queue.
Another approach is to find an independent set for the first partite set, then repeat for a second partite set, as in Luby's Algorithm~\cite{luby1986simple}. Recent work showed that by using the minimum-degree heuristic for independent set, this strategy gives a $d$-approximation in $d$-degenerate graphs~\cite{goodrich2017optimizing}. Both of these methods are nondeterministic; the former is stochastic
by design, and the latter breaks ties between minimum degree vertices.
We provide C++ implementations of these four heuristics (\dfs, \bfs, \luby, and \mindeg), and provide a \emph{heuristic ensemble} solver that runs the heuristics round-robin until a specified timeout is reached.

\subsection{Iterative Compression}
\label{subsection:huffner}

The state-of-the-art implementation for solving \myoct combinatorially comes from \huffner's simplification of the iterative compression algorithm \cite{huffner2009algorithm, reed2004finding}.
Broadly, the iterative compression technique starts with a trivial solution on a subgraph of the instance, expands both the solution and subgraph, then applies a compression routine to reduce the solution if possible.
By iterating this process, the subgraph eventually encompasses the full graph, and the solution at every step is compressed to remain within some desired bound.
The compression routine may have run time exponential in the size of the solution and is applied at most a linear number of times, naturally leading to an FPT algorithm parameterized by the solution size $k$.

In the specific application to \myoct, the compression routine tries all $O(3^k)$ partitions of a $(k+1)$-sized solution into a new transversal and left/right partite sets.
For each partition, an \myoct set for the full subgraph is computed by solving a min-cut instance.
If the number of vertices assigned to the transversal plus the vertices removed by the cut is less than $k+1$, then the solution was compressed, and otherwise a certificate is found that no solution of size $k$ exists on this subgraph.
Using Edmonds-Karp for the min-cut algorithm, this compression routine runs in $O(3^k \cdot k \cdot |E|)$.

The iterating outer loop that expands the solution and subgraph is trivial for \myoct: Given an ordering of the vertices, the initial subgraph and solution are the first $k$ vertices, and the subgraph and solution are both expanded by adding the next unused vertex in the ordering to each.
In the worst case there are $n$ iterations, resulting in a total run time of $O(3^k \cdot k \cdot |V| \cdot |E|)$.

\subsection{Modifications to \huffner's Algorithm}
While \huffner's reformulation of Reed et al.'s algorithm was based on improving the compression routine, we note some straightforward improvements to the outer loop that can also lead to improvements in practice.
These improvements are related to the choice of vertex ordering.

First, the number of compressions can be reduced by choosing an initial subgraph larger than the initial solution of size $k$.
Specifically, given a heuristic solution $S$ for \myoct, we can construct an initial subgraph of size $\min(|V|-|S|+k, |V|)$ by placing the vertices in $V \setminus S$ first in the ordering, then initializing the subgraph with this bipartite subgraph and up to $k$ of the remaining vertices.
This `bipartite jump-start' has no negative effect on the theoretical run time, but may improve run time by up to a factor of $|V|$ based on the quality of the heuristic solution.

Second, \huffner's improvements to the compression routine utilize the presence of edges to eliminate possible partitions from consideration.
Namely, two vertices cannot be assigned to the same partition if they share an edge, and the number of partitions can be reduced from $O(3^k)$ in the worst case (an independent set) to $O(k^2)$ in the best case (a clique).
To exploit this fact, after the ordering is jump-started with a bipartite subgraph, we order the remaining vertices in a reverse degeneracy ordering~\cite{lick1970k}.
This ordering guarantees that vertices added to the subgraph maximize the number of newly introduced edges.

Finally, we note that iterative compression is naturally an anytime algorithm.
If, at any point, the iteration stops, the current solution size is a lower bound on the optimal \myoct, and the current solution plus the vertices not yet reached in the ordering form an upper bound.
By its FPT nature, the iterative compression approach becomes hard when $k$ becomes large.
However, iterative compression may fill an important niche between heuristics and exact solvers by offering a structured approach for compressing heuristic solutions as time allows.

We provide a modified copy of \huffner's implementation with both improvements implemented in C++ and enable anytime functionality when given a termination signal.
Based on small-scale experimentation (Appendix \ref{section:extended_ic}), we find that the bipartite jump-start always helps in expectation, but the degeneracy ordering may not be worth the additional run time when using a small timeout.
In our experiments we use only the first improvement on timeouts of 0.01 and 0.1 seconds, and otherwise we use both improvements.

\subsection{Alternative Solvers}
\label{section:alternative_solvers}
As mentioned in Section \ref{section:background}, \myoct can be converted into an equivalent \vcfull (\vc) or \ilpfull (\ilp) instance.

Currently, the fastest theoretical run time for \myoct and other related problems comes from a \vc-formulation~\cite{lokshtanov2014faster}.
We evaluate this approach with Akiba and Iwata's Java solver, which was previously demonstrated to be faster than \huffner's iterative compression algorithm \cite{akiba2016branch}.
We implement a Python wrapper that converts between \myoct and \vc, providing a common API with the other solvers.

When solving \myoct as an \ilp, the instance can be directly converted $\myoct \to \ilp$, or converted with $\myoct \to \vc \to \ilp$.
In addition to choosing a formulation, the user must also choose an \ilp solver (e.g. \textsf{CPLEX} or \textsf{GLPK}) and consider the effect of additional threads and/or RAM limitations.
Testing several configurations (Appendix~\ref{section:extended_ilp}), we confirmed several best practices from previous work.
The biggest factor was choice of solver, where IBM's closed-source \cplex solver bested the open-source GNU solver \glpk.
The next factor with the biggest impact was choice of formulation, where $\myoct \to \vc \to \ilp$ performed significantly better than $\myoct \to \ilp$; this performance difference may be explained by a similar result in theoretical analysis \cite{lokshtanov2014faster}.

When evaluating the scalability of \cplex on server hardware, we found that using multiple threads may lead to super-linear speed-up, and that while RAM limitations may increase run time, they do not change the relationship between other factors.
In our experiments, we use the $\myoct \to \vc \to \ilp$ formulation with the \cplex solver, a single thread, and unrestricted RAM.
Additionally, \cplex allows recovery of partial solutions, enabling us to use this approach as an anytime algorithm.
We again provide a Python wrapper for a common API.

\section{Data Benchmark and Code}
\label{section:data_benchmark_and_code}
In this section we detail the data used in the experiments, along with the code made available at \cite{practical-oct}.

\subsection{Previous Data}
\label{subsection:data}
As mentioned in Section~\ref{subsection:related_work}, the primary dataset studied in \myoct literature originates from Wernicke and is distributed with \huffner's code.
We refer to this data as the \texttt{Wernicke-\huffner} (\texttt{WH}) dataset.
These datasets are originally from genetics through the related \textsc{Minimum Site Removal} problem \cite{wernicke2003algorithmic} and are expected to be close to bipartite.
This dataset is composed of 45 Afro-American graphs (denoted \texttt{aa-10}, $\dots$, \texttt{aa-54}) and 29 Japanese graphs (denoted \texttt{j-10}, $\dots$, \texttt{j-28}).
Files \texttt{aa-12}, \texttt{j-12}, and \texttt{j-27} are provided in \huffner's code, but are empty and excluded here.

\subsection{Quantum-Inspired Data}

While the \wh dataset may have been of historical interest, recent results show that any state-of-the-art solver can
solve all these instances within three seconds \cite{akiba2016branch}. To introduce a new benchmark corpus for \myoct
solvers, we concentrate on domain data from quantum computing. Specifically, a recent survey \cite{dunning2015works}
collected six datasets from the \qubo literature (see Section~\ref{section:background}); of these, only the \texttt{Beasley}
\cite{beasley1998heuristic} and \texttt{GKA} \cite{glover1998adaptive} data have instances small enough to embed in
near-term quantum annealing hardware.

In this work we consider the 50-vertex instances (denoted \texttt{b-50-1}, $\dots$, \texttt{b-50-10}) and the 100-vertex
instances (denoted \texttt{b-100-1}, $\dots$, \texttt{b-100-10}) from the \texttt{Beasley} dataset and the first 35 instances
of the \texttt{GKA} dataset (denoted \texttt{gka-1}, $\dots$, \texttt{gka-35}), which have varying numbers of vertices
and densities (c.f., Table~\ref{table:preprocessed_summary}). All \qubo datasets are parsed as undirected, simple graphs
with no vertex or edge weights. Vertices, edges with weight zero, and self-loops are excluded.

Additionally, we examine frustrated cluster loop (FCL) graphs. FCLs are generated by overlaying cycles from random walks
on a base graph and were originally introduced as a benchmark for quantum annealing hardware that produced \qubo instances
with \emph{rugged} energy landscapes \cite{king2019quantum}. Rugged landscapes are characterized by having frequent, tall
energy barriers, which reduces the effectiveness of classical solvers and makes use of quantum tunneling.

We use the D-Wave NetworkX API to generate a corpus of FCL instances applicable to four hardware models: a D-Wave 2000Q
($16 \times 16$ Chimera grid), a theoretical D-Wave ``8000Q'' ($32 \times 32$ Chimera grid), the upcoming Pegasus-6
(which embeds a $K_{52,52}$), and a Pegasus-12 (which embeds a $K_{124, 124}$). For each hardware, we generated a corpus
of FCLs from a clique of size $n$ for a subset of the values between the size of the maximum clique embedding of the hardware
and the size of the maximum biclique embedding of the hardware. See Appendix~\ref{appendix:datasets} for step sizes.
We also varied the number of cycles used in each FCL from 5-30\% of the base clique size to ensure a diverse set of FCL
instances in the regime where \myoct-embedding methods would be deployed. For example, the D-Wave 2000Q can embed a
64-clique and a 128-biclique, so we generate FCLs using cliques of size 64, 80, 96, 112, and 128; as seen in
Figure~\ref{figure:fcl} in Appendix~\ref{appendix:datasets}, the varied number of cycles results in a corpus of
instances with good coverage of the desired size interval $[64, 128]$.

Due to similarity between the graph instances sizes, we further cluster the 2000Q and Pegasus-6
FCL graphs into a \texttt{fcl-small} corpus and the 8000Q and Pegasus-12 into \texttt{fcl-large}.

\subsection{Synthetic Graph Generators}
\label{section:synthetic_generators}
To generalize our results and prevent bias that may be present in a difficult \qubo benchmark, we use synthetic graph generators to mimic distinct properties of the quantum graphs.

To match edge density, we use the \er model \cite{erdos1960evolution}, which takes as input a number of vertices $n$ and a probability $p$.
\er generates a graph by initializing $n$ vertices and adding each edge with probability $p$.
By setting $p := |E| / \binom{n}{2}$ we have the same edge density in expectation.

To mimic a dataset's distance-to-bipartite, we provide a \tunoct generator as a modification of \er.
The \tunoct generator requires an upper bound on the optimal \myoct solution (denoted $n_o$) and a bipartite balance parameter $0 \leq b \leq 1$.
\tunoct generates a graph by partitioning the vertices into an odd cycle transversal, a left partite set, and a right partite set.
The odd cycle transversal has $n_o$ vertices, and the remaining vertices are assigned to the left partite set with probability $b$.
Edges are then generated according to \er, with the exception that vertices in the same partite set may not share an edge.
This construction enforces that $OPT \leq n_o$, highlighting the distinction between arbitrary edge placements and a potentially small optimal \myoct solution.
In our experimental results we set $n_o$ equal to the optimal solution for the original (non-preprocessed) graphs, and set $b = 0.5$.

For matching degree distribution in addition to density we use the \cl expected degree model \cite{chung2002connected}.
Given a degree distribution $(d_1, \dots, d_n)$, the \cl model adds an edge $uv$ with probability
\begin{align*}
P_{uv} = \frac{d_u d_v}{\sum_{i=1}^n d_i}.
\end{align*}
We generate these graphs using the original instances' degree distribution.

Finally, we also include the \ba preferential attachment model \cite{albert2002statistical} to highlight the effect of a severely biased degree distribution at fixed edge density.
Given a set of initial vertices, a constant $c$, and a number $n$ of additional vertices, each new vertex is added to the graph with $c$ edges attached to the existing nodes with probability proportional to their current degree.
We match the original graphs' number of vertices and select $c$ such that the same number of edges are added (up to integer rounding).

\subsection{Replicability}
All experiments are fully replicable with our open source code repository~\cite{practical-oct}.
After installing the software with the \texttt{README} instructions, we direct the interested reader to \replicability for detailed instructions.

To sanitize the data, graphs are relabeled with vertices $\{0, \dots, n-1\}$ and are written to files for each solver's required format.
The reduction routines are natively nondeterministic, but data is explicitly sorted in both \myoct- and \vc-based routines such that a single run will generate identical results on distinct hardware and software environments.

All algorithms are available as standalone solvers using a command-line interface, and Python scripts are provided for reproducing the experiments, including tables and plots.

\section{Quantum-Specific Results}
\label{section:quantum_results}
\subsection{Experimental Setup}
All experiments were run on three identical Dell PowerEdge R430 servers, each
with an Intel E5-2623 v2 CPU (3.5GHz single-core turbo, 10MB cache) and 64GB ECC DDR4 RAM.
Each server ran Fedora 27 with Linux kernel 4.16.7, CPLEX 12.8, and GLPK 4.61.
C and C++ code were compiled with Clang version 5.0.1, Java code was compiled with OpenJDK 1.8.0\_181, and Python code was run with Python 3.5.6 (restricted by CPLEX 12.8).

\subsection{Preprocessing Effectiveness}
\label{subsection:preprocessing}

\begin{table*}[!t]
\tabcolsep = 1.5mm
\begin{tabular}{lcc c ccccc c cc}
\toprule
& \multicolumn{2}{c}{\textbf{Original Graph}} & \phantom{00} & \multicolumn{5}{c}{\textbf{Reductions}} & \phantom{00} & \multicolumn{2}{c}{\textbf{Reduced Graph}} \\
\cmidrule(lr){2-3}
\cmidrule(lr){5-9}
\cmidrule(lr){11-12}
\textbf{Dataset} &    $|V|$ &  $|E|/|V|$ && $|\widehat{V_r}|$ & $|\widehat{E_r}|$ & $|\widehat{V_o}|$ & $|\widehat{V_b}|$ & \textbf{Solved} && $|V' \cup V_b|$ & $|E'|/|V' \cup V_b|$ \\
\midrule
\texttt{WH-aa} &
39--300 &
1.8--5.4 &&
13\% &
- &
- &
70\% &
11\% &&
27--265 &
1.9--6.1 \\
\texttt{WH-j} &
33--150 &
1.5--5.9 &&
29\% &
- &
- &
13\% &
23\% &&
8--74 &
2.0--7.5 \\
\texttt{b-50} &
50 &
2.0--2.7 &&
2\% &
- &
- &
4\% &
- &&
43--50 &
2.1--2.7 \\
\texttt{b-100} &
100 &
4.6--5.1 &&
- &
- &
- &
- &
- &&
100 &
4.6--5.1 \\
\texttt{gka} &
20--125 &
2.1--61.3 &&
- &
- &
9\% &
- &
11\% &&
6--100 &
2.0--43.5 \\
\bottomrule
\end{tabular}

\caption{A summary of preprocessing statistics on \wh and quantum datasets. Ranges are given for the number of vertices and edge density in both the original and reduced graphs.
The statistic $|\widehat{V_r}|$ reports the percentage of vertices removed on average. Likewise, normalized means are reported for edge removals $E_r$, fixed-OCT vertices $V_o$, and fixed-bipartite vertices $V_b$; dashes denote zero changes. The percentage of graphs solved completely by preprocessing routines is also reported.}
\label{table:preprocessed_summary}
\end{table*}

In this subsection we harden the benchmark by applying the reduction routines detailed in Section~\ref{section:reduction_routines}.
We denote reductions as a partition of the original vertex set $V = V_r \uplus V_o \uplus V_b \uplus V'$.
The vertices that may be removed without changing \myoct are denoted $V_r$. For some fixed optimal solution $S$, vertices $V_o$ must be in $S$ and vertices $V_b$ cannot be in $S$.
Finally, the remaining vertices are labeled $V'$.
Analogously, the edges are partitioned into $E = E_r \uplus E'$.
The sets $V_r$, $E_r$, and $V_o$ may be safely removed from the graph, leaving the reduced graph with vertices $V_b \cup V'$ and edges $E'$.

Table~\ref{table:preprocessed_summary} summarizes preprocessing results over the non-synthetic datasets.
We observe that the reduction routines' effectiveness is dataset-dependent.
The \wh data is amenable to vertex removals, particularly \texttt{WH-j}, which had its largest graph reduced from 241 to 74 vertices.
Perhaps due to a very low edge density, the \texttt{WH-aa} also had a significant number of vertices labeled bipartite.
The \gka dataset was only affected by reductions that fixed \myoct vertices, which can be important for exact algorithms fixed-parameter tractable in the solution size.
Few reductions applied to the \beasley data, with \texttt{b-100} remaining untouched.
Both \wh and \gka data contained instances that were completely solved by preprocessing.

\subsection{Use Case: Heuristic Solutions}
For the first quantum use case, an embedding compiler may need a bipartization of an input program (e.g., a \textsc{QUBO} or circuit) in order to prune a search space of embeddings, but has very little time budgeted for this step.
Therefore we compare the best (potentially non-optimal) solutions found per solver with fixed timeouts.
The algorithms capable of producing heuristic solutions are the heuristic ensemble (\textsf{HE}), the improved iterative compression solver (\textsf{IC}), and the \ilp formulation with \cplex (\textsf{ILP}).
To evaluate the heuristics and anytime algorithms in this scenario we choose timeouts of four different orders of magnitude (0.01, 0.1, 1.0, and 10 seconds).
At the timeout, each algorithm is given a termination signal and is given time to output the last solution cached.

Table~\ref{table:heuristic-approx} reports the worst-case approximation factors achieved by an algorithm-data-timeout triple; full results on the \wh, \beasley, and \gka data are reported in Appendix~\ref{appendix:extended_results} (Tables~\ref{table:heuristic1} and \ref{table:heuristic2}).
We observe that for very short timeouts, the \textsf{HE} and \textsf{IC} solvers noticeably outperform the \textsf{ILP} solver in terms of solution quality, achieving worst-case approximation factors of at most $1.57$ for all data with timeouts of 0.01(s) and 0.1(s). At the other end of the spectrum, for a timeout of 10s, \textsf{ILP} is the dominant solver with a worst-case approximation factor of $1.21$. Between these two at a timeout of 1s, the best approach is not clear. We also note that the approximation factors for the FCL datasets are consistently larger than those of the non-synthetic data.

\begin{table*}[t]
\tabcolsep = 1.3mm
\begin{tabular}{lc c ccc c ccc c ccc c ccc}
\toprule
\multicolumn{2}{r}{\textbf{Timeout:}} &
\multicolumn{3}{c}{\textbf{0.01(s)}} & \phantom{Q} &
\multicolumn{3}{c}{\textbf{0.1(s)}} & \phantom{Q} &
\multicolumn{3}{c}{\textbf{1(s)}} & \phantom{Q} &
\multicolumn{3}{c}{\textbf{10(s)}}\\
\cmidrule(lr){3-5}
\cmidrule(lr){7-9}
\cmidrule(lr){11-13}
\cmidrule(lr){15-17}
\textbf{Dataset} &&
\textsf{HE} & \textsf{IC} & \textsf{ILP} &&
\textsf{HE} & \textsf{IC} & \textsf{ILP} &&
\textsf{HE} & \textsf{IC} & \textsf{ILP} &&
\textsf{HE} & \textsf{IC} & \textsf{ILP}\\
\midrule

\texttt{WH-aa} &&
\textbf{1.54} & 1.57 & 24.25 &&
1.33 & \textbf{1.24} & 2.10 &&
1.29 & 1.29 & \textbf{1.25} &&
1.24 & 1.08 & \checkmark\\

\texttt{WH-j} &&
1.11 & \checkmark & 3.92 &&
1.11 & \checkmark & \checkmark &&
\checkmark & \checkmark & \checkmark &&
\checkmark & \checkmark & \checkmark\\

\texttt{b-50} &&
1.11 & \textbf{1.09} & 1.67 &&
1.11 & \checkmark & 1.21 &&
1.08 & \checkmark & \checkmark &&
\checkmark & \checkmark & \checkmark\\

\texttt{b-100} &&
\textbf{1.07} & 1.10 & 2.10 &&
\textbf{1.07} & 1.10 & 1.34 &&
\textbf{1.05} & 1.07 & 1.15 &&
\textbf{1.05} & \textbf{1.05} & \textbf{1.05}\\

\texttt{gka} &&
\textbf{1.18} & 1.21 & 2.36 &&
\textbf{1.18} & 1.21 & 1.36 &&
\textbf{1.11} & \textbf{1.11} & 1.17 &&
1.11 & 1.11 & \textbf{1.07}\\

\texttt{fcl-small} &&
1.50 & \textbf{1.41} & 5.00 &&
1.50 & \textbf{1.29} & 1.57 &&
1.33 & 1.20 & \textbf{1.18} &&
1.23 & 1.18 & \textbf{1.09}\\

\texttt{fcl-large} &&
1.71 & \textbf{1.50} & 5.00 &&
1.53 & \textbf{1.42} & 2.33 &&
1.43 & \textbf{1.35} & 1.48 &&
1.38 & 1.29 & \textbf{1.21}\\
\bottomrule
\end{tabular}

\caption{Observed approximation factors for anytime algorithms: the heuristic ensemble (\textsf{HE}), iterative compression (\textsf{IC}), and integer linear programming (\textsf{ILP}). For each dataset, the worst-case approximation ratio over its instances is reported. Approximation ratios are with respect to \myoct on the reduced graph, computed with \textsf{ILP}. A checkmark denotes that exact solutions are found on all instances, if a dataset has no checkmark then the best approximation algorithm is bolded.}
\label{table:heuristic-approx}
\end{table*}

\begin{table*}[!t]
\tabcolsep = 1.5mm
\begin{tabular}{lrrrrp{0.5in}lrrrr}
\toprule
  \multicolumn{2}{c}{\textbf{Graph}}            & \multicolumn{3}{c}{\textbf{Solver}}            &            & \multicolumn{2}{c}{\textbf{Graph}}            & \multicolumn{3}{c}{\textbf{Solver}} \\
\cmidrule(lr){1-2}
\cmidrule(lr){3-5}
\cmidrule(lr){7-8}
\cmidrule(lr){9-11}
\textbf{Dataset} & $OPT$ & \textsf{VC} & \textsf{IC} & \textsf{ILP} & & \textbf{Dataset} & $OPT$ & \textsf{VC} & \textsf{IC} & \textsf{ILP} \\
\midrule
\texttt{b-100-1} & 41 & 101.5 & - & \textbf{79.5} && \texttt{gka-21} & 40 & 1.5 & 31.2 & \textbf{0.6} \\
\texttt{b-100-2} & 42 & 190.8 & - & \textbf{159.2} && \texttt{gka-22} & 43 & \textbf{5.0} & - & 6.5 \\
\texttt{b-100-3} & 42 & 252.8 & - & \textbf{58.0} && \texttt{gka-23} & 46 & \textbf{44.5} & - & 63.2 \\
\texttt{b-100-4} & 41 & 212.7 & - & \textbf{176.8} && \texttt{gka-24} & 37 & \textbf{76.6} & - & 95.8 \\
\texttt{b-100-5} & 42 & \textbf{217.6} & - & 246 && \texttt{gka-25} & 42 & \textbf{130.6} & - & 169.6 \\
\texttt{b-100-6} & 43 & \textbf{150.1} & - & 223.3 && \texttt{gka-26} & 43 & 168.2 & - & \textbf{113.6} \\
\texttt{b-100-7} & 42 & \textbf{189.0} & - & 270.1 && \texttt{gka-27} & 62 & 555.1 & - & \textbf{477.3} \\
\texttt{b-100-8} & 43 & 368.7 & - & \textbf{209.6} && \texttt{gka-28} & 70 & \textbf{300.8} & - & 516.1 \\
\texttt{b-100-9} & 44 & \textbf{333.5} & - & 470.5 && \texttt{gka-29} & 77 & \textbf{67.1} & - & 423.0 \\
\texttt{b-100-10} & 44 & \textbf{195.4} & - & 478.4 && \texttt{gka-30} & 82 & \textbf{33.6} & - & 305.2 \\
\texttt{gka-3} & 23 & \textbf{1.5} & 273.4 & 4.4 && \texttt{gka-31} & 85 & \textbf{12.4} & 109.6 & 91.9 \\
\texttt{gka-8} & 28 & \textbf{7.2} & - & 8.5 && \texttt{gka-32} & 88 & \textbf{5.4} & 18.8 & 40.3 \\
\bottomrule
\end{tabular}

\caption{Run times (in seconds) of exact solvers on a representative sample of \beasley and \gka data with a 10 minute timeout. Algorithm-data pairings that did not finish within the timeout are denoted with a dash, and the best run time on a dataset is bolded.}
\label{table:exact_summary}
\end{table*}

\subsection{Use Case: Exact Solutions}
In the second use case, a researcher may want a `litmus test' evaluating whether a program's structure can be represented in a given quantum hardware.
By recognizing that the \myoct size of a problem graph is too large, we can show that such a configuration is impossible to embed in this hardware topology.
We simulate this use case by computing exact solutions with a 10 minute timeout on preprocessed instances using the three solvers that could guarantee optimality: the \vc-solver (\textsf{VC}), iterative compression (\textsf{IC}), and \ilp (\textsf{ILP}).

We find that the solvers are fairly evenly split on total run time victories (Table~\ref{table:exact_summary}),  with \textsf{ILP} performing better on the \beasley data and \textsf{VC} performing better on \gka.
Iterative compression only finished a handful of times within 10 minutes, and only once was competitive (\texttt{gka-32}).

\begin{figure}
\includegraphics[width=0.6\textwidth]{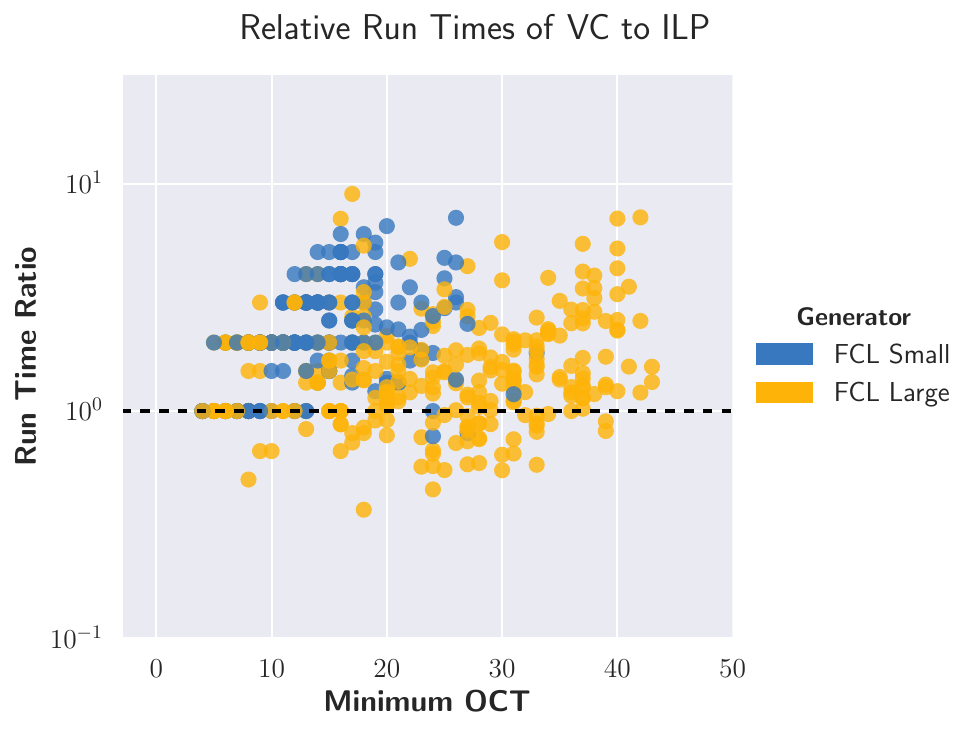}
\caption{Plot of \vc-solver time divided by \ilp solver time for exact solutions on FCL graphs. Points above the dashed line indicate the ILP formulation outperforms the equivalent VC formulation, and include almost all of the \texttt{fcl-small} instances.}
\label{fig:fcl-exact-ratio}
\end{figure}

Based on these results, we chose to only run the \vc-solver and \ilp on the FCL data. As seen in Figure~\ref{fig:fcl-exact-ratio}, the \ilp solver was dominant. While the \vc-solver was faster on a few \texttt{fcl-large} instances, this does not appear to be correlated with the minimum OCT size.

\section{Generalized Results}
\label{section:generalized_results}
In this section, we expand the experimental envelope further by taking each \wh and quantum dataset and creating several ``look-a-likes'' -- synthetic graphs that match the original in different facets.
We use the \er generator to match density, the \tunoct model to match proximity to bipartite, the \cl model to match degree distribution, and \ba to increase degree distribution heterogeneity at a fixed edge density; all models and parameter settings are detailed in Section~\ref{section:synthetic_generators}.
A single synthetic instance is generated from a quantum instance, a synthetic graph generator, and a pseudorandom number generator seed.
15 seeds are used for the preprocessing and heuristic experiments, and 5 seeds are used for the (computationally expensive) exact results.

\subsection{Reduction Effectiveness}
Table~\ref{table:preprocessed_synthetic_summary} depicts the effectiveness of reduction routines on the synthetic data.
Across most datasets, the \tunoct and \cl synthetic graphs resulted in more vertex removals than the original graphs.
In the case of \texttt{WH-aa}, this increase in $V_r$ also involved a drastic decrease of vertices labeled bipartite ($V_b$).
Again, the \beasley datasets remained relatively unaffected by reduction, excepting the \tunoct analogues for \texttt{b-50}.

\begin{table*}[t]
\tabcolsep = 1.5mm
\begin{tabular}{lcc c ccccc c cc}
\toprule
& \multicolumn{2}{c}{\textbf{Original Graph}} & \phantom{00} & \multicolumn{5}{c}{\textbf{Reductions}} & \phantom{00} & \multicolumn{2}{c}{\textbf{Reduced Graph}} \\
\cmidrule(lr){2-3}
\cmidrule(lr){5-9}
\cmidrule(lr){11-12}
\textbf{Dataset} &    $|V|$ &  $|E|/|V|$ && $|\widehat{V_r}|$ & $|\widehat{E_r}|$ & $|\widehat{V_o}|$ & $|\widehat{V_b}|$ & \textbf{Solved} && $|V' \cup V_b|$ & $|E'|/|V' \cup V_b|$ \\
\midrule
\textbf{\texttt{WH-aa}} &
\textbf{39--300} &
\textbf{1.8--5.4} &&
\textbf{13\%} &
\textbf{-} &
\textbf{-} &
\textbf{70\%} &
\textbf{11\%} &&
\textbf{27--265} &
\textbf{1.9--6.1} \\
\midrule
\texttt{WH-aa-ER} &  15--300 &   1.2--5.6 &&               1\% &                 - &                 - &               2\% &             4\% &&         10--300 &             1.4--5.6 \\
\texttt{WH-aa-TO} &  27--300 &   0.9--4.1 &&              15\% &                 - &                 - &              12\% &             8\% &&         10--299 &             1.4--4.1 \\
\texttt{WH-aa-CL} &  27--300 &   1.3--5.5 &&              55\% &                 - &               6\% &              12\% &             9\% &&          9--155 &             1.5--6.7 \\
\texttt{WH-aa-BA} &  14--300 &   1.7--5.9 &&               1\% &                 - &                 - &               1\% &             8\% &&          7--300 &             1.4--5.9 \\
\midrule
\textbf{\texttt{WH-j}} &
\textbf{33--150} &
\textbf{1.5--5.9} &&
\textbf{29\%} &
\textbf{-} &
\textbf{-} &
\textbf{13\%} &
\textbf{23\%} &&
\textbf{8--74} &
\textbf{2.0--7.5} \\
\midrule
\texttt{WH-j-ER} &  33--241 &   1.1--6.3 &&               6\% &                 - &                 - &               4\% &            13\% &&         18--239 &             1.5--6.3 \\
\texttt{WH-j-TO} &  33--150 &   0.8--4.1 &&              17\% &                 - &                 - &               7\% &            32\% &&         10--104 &             1.4--4.1 \\
\texttt{WH-j-CL} &  33--241 &   1.2--6.2 &&              39\% &                 - &               4\% &               7\% &            22\% &&         10--138 &             1.5--7.5 \\
\texttt{WH-j-BA} &  33--241 &   1.9--5.6 &&               7\% &                 - &               1\% &               4\% &            14\% &&         10--241 &             1.3--5.7 \\
\midrule
\textbf{\texttt{b-50}} &
\textbf{50} &
\textbf{2.0--2.7} &&
\textbf{2\%} &
\textbf{-} &
\textbf{-} &
\textbf{4\%} &
\textbf{-} &&
\textbf{43--50} &
\textbf{2.1--2.7} \\
\midrule
\texttt{b-50-ER} &       50 &   1.7--3.3 &&               3\% &                 - &                 - &               3\% &               - &&          31--50 &             1.9--3.4 \\
\texttt{b-50-TO} &       50 &   1.2--2.5 &&              14\% &                 - &                 - &               6\% &             3\% &&          13--48 &             1.4--2.7 \\
\texttt{b-50-CL} &       50 &   1.7--3.0 &&               8\% &                 - &               1\% &               4\% &               - &&          24--49 &             1.8--3.2 \\
\texttt{b-50-BA} &       50 &        2.8 &&               4\% &                 - &               2\% &               5\% &             6\% &&          30--50 &             1.9--2.8 \\
\midrule
\textbf{\texttt{b-100}} &
\textbf{100} &
\textbf{4.6--5.1} &&
\textbf{-} &
\textbf{-} &
\textbf{-} &
\textbf{-} &
\textbf{-} &&
\textbf{100} &
\textbf{4.6--5.1} \\
\midrule
\texttt{b-100-ER} &      100 &   4.3--5.5 &&                 - &                 - &                 - &                 - &               - &&         98--100 &             4.3--5.5 \\
\texttt{b-100-TO} &      100 &   3.6--4.8 &&                 - &                 - &                 - &               1\% &               - &&         95--100 &             3.6--4.8 \\
\texttt{b-100-CL} &      100 &   4.1--5.5 &&               1\% &                 - &                 - &               1\% &               - &&         95--100 &             4.2--5.5 \\
\texttt{b-100-BA} &      100 &   4.8--5.6 &&                 - &                 - &                 - &                 - &               - &&         98--100 &             4.3--5.7 \\
\midrule
\textbf{\texttt{gka}} &
\textbf{20--125} &
\textbf{2.1--61.3} &&
\textbf{-} &
\textbf{-} &
\textbf{9\%} &
\textbf{-} &
\textbf{11\%} &&
\textbf{6--100} &
\textbf{2.0--43.5} \\
\midrule
\texttt{gka-ER} &  20--125 &  1.8--61.4 &&                 - &                 - &               9\% &                 - &            13\% &&          6--100 &            1.9--43.7 \\
\texttt{gka-TO} &  20--125 &  1.2--61.4 &&               1\% &                 - &              12\% &               1\% &             8\% &&          6--100 &            1.6--43.6 \\
\texttt{gka-CL} &  20--125 &  1.7--60.9 &&               1\% &                 - &              10\% &                 - &            11\% &&          6--100 &            1.7--43.2 \\
\texttt{gka-BA} &  20--125 &  2.8--31.2 &&                 - &                 - &               1\% &                 - &             1\% &&          9--123 &            1.6--29.9 \\
\bottomrule
\end{tabular}

\caption{A summary of preprocessing statistics over all datasets. Ranges are given for the number of vertices and edge density in both the original and reduced graphs.
The normalized statistic $|\widehat{V_r}|$ reports the average percentage of vertices removed. Likewise, normalized means are reported for edge removals $E_r$, fixed-OCT vertices $V_o$, and fixed-bipartite vertices $V_b$; dashes denote zero changes.
The percent of graphs per dataset completely solved by preprocessing routines is also noted. Results are reported over 15 seeds per random graph generator, per original dataset.}
\label{table:preprocessed_synthetic_summary}
\end{table*}

\subsection{Heuristic Solutions}
When extending the heuristic solution comparison to synthetic data, we find the same best practices generally still apply (Table~\ref{table:heuristic-approx-synthetic}).
At 0.01s, \textsf{IC} has little time to improve on the initial heuristic solution and so the quick solutions found by \textsf{HE} remain competitive.
At 0.1s and 1s, \textsf{HE} begins to have diminishing returns and the compression steps towards an exact solution begin to pay off for \textsf{IC}.
At 1s, \textsf{ILP} also begins to produce winning solutions, solving all of \texttt{b-50} and splitting the wins otherwise.
Finally, with 10s to execute the full CPLEX reduction routines, \textsf{ILP} begins to dominate.

We observe that over all data and algorithms, 0.01 seconds is enough to get a 2.70-approximation (and 1.42 outside of \texttt{WH-aa-to}) with \textsf{IC}; 0.1 seconds yields a 1.79-approximation with \textsf{IC}; 1 second yields a 1.65-approximation with \textsf{IC}, and 10 seconds yields a 1.14-approximation with \textsf{ILP}.
From a domain-science standpoint, these relatively small errors show that \myoct can be considered a practical subroutine when searching for a `good enough' solution in a time-sensitive environment.

\begin{table*}[!t]
\tabcolsep = 1.3mm
\begin{tabular}{lc c ccc c ccc c ccc c ccc}
\toprule
\multicolumn{3}{r}{\textbf{Timeout:}} &
\multicolumn{3}{c}{\textbf{0.01(s)}} & \phantom{Q} &
\multicolumn{3}{c}{\textbf{0.1(s)}} & \phantom{Q} &
\multicolumn{3}{c}{\textbf{1(s)}} & \phantom{Q} &
\multicolumn{3}{c}{\textbf{10(s)}}\\
\cmidrule(lr){4-6}
\cmidrule(lr){8-10}
\cmidrule(lr){12-14}
\cmidrule(lr){16-18}
\textbf{Dataset} & \textbf{Represented} &&
\textsf{HE} & \textsf{IC} & \textsf{ILP} &&
\textsf{HE} & \textsf{IC} & \textsf{ILP} &&
\textsf{HE} & \textsf{IC} & \textsf{ILP} &&
\textsf{HE} & \textsf{IC} & \textsf{ILP}\\
\midrule
\texttt{WH-aa-er} & 61\%&& \textbf{1.26} & 1.30 & 3.16&& \textbf{1.21} & \textbf{1.21} & 1.52&& 1.21 & \textbf{1.16} & 1.35&& \textbf{1.14} & \textbf{1.14} & 1.15\\
\texttt{WH-aa-to} & 100\%&& \textbf{2.41} & 2.70 & 15.00&& 1.95 & \textbf{1.79} & 5.75&& \textbf{1.65} & \textbf{1.65} & 3.11&& 1.54 & 1.56 & \checkmark\\
\texttt{WH-aa-cl} & 100\%&& 1.33 & \textbf{1.24} & 3.43&& 1.25 & \textbf{1.21} & 1.47&& 1.18 & 1.19 & \textbf{1.11}&& 1.17 & 1.17 & \textbf{1.05}\\
\texttt{WH-aa-ba} & 86\%&& 1.50 & \textbf{1.39} & 4.43&& 1.40 & \textbf{1.33} & 1.73&& 1.33 & \textbf{1.30} & 1.39&& 1.30 & 1.30 & \textbf{1.14}\\
\midrule
\texttt{WH-j-er} & 93\%&& \textbf{1.30} & 1.31 & 3.12&& \textbf{1.21} & \textbf{1.21} & 1.64&& 1.19 & \textbf{1.18} & 1.39&& 1.19 & 1.18 & \textbf{1.11}\\
\texttt{WH-j-to} & 100\%&& 1.78 & \textbf{1.31} & 8.33&& 1.44 & \textbf{1.18} & 1.88&& 1.31 & 1.11 & \checkmark&& 1.15 & \checkmark & \checkmark\\
\texttt{WH-j-cl} & 100\%&& \textbf{1.29} & \textbf{1.29} & 3.67&& 1.25 & \textbf{1.17} & 1.52&& 1.21 & 1.17 & \textbf{1.08}&& 1.17 & 1.15 & \checkmark\\
\texttt{WH-j-ba} & 100\%&& 1.60 & \textbf{1.42} & 4.58&& 1.40 & \textbf{1.33} & 1.93&& \textbf{1.33} & 1.36 & 1.48&& 1.31 & 1.36 & \textbf{1.10}\\
\midrule
\texttt{b-50-er} & 100\%&& 1.33 & \textbf{1.17} & 2.00&& 1.18 & \textbf{1.08} & 1.31&& 1.12 & 1.07 & \checkmark&& \checkmark & 1.07 & \checkmark\\
\texttt{b-50-to} & 100\%&& 1.25 & \textbf{1.10} & 2.25&& 1.20 & \textbf{1.14} & 1.22&& \checkmark & \checkmark & \checkmark&& \checkmark & \checkmark & \checkmark\\
\texttt{b-50-cl} & 100\%&& 1.20 & \textbf{1.17} & 1.75&& 1.17 & \textbf{1.12} & 1.27&& 1.14 & 1.08 & \checkmark&& \checkmark & 1.08 & \checkmark\\
\texttt{b-50-ba} & 100\%&& 1.25 & \textbf{1.18} & 1.90&& 1.17 & \textbf{1.09} & 1.18&& 1.10 & 1.09 & \checkmark&& \checkmark & \checkmark & \checkmark\\
\midrule
\texttt{b-100-er} & 100\%&& \textbf{1.10} & 1.19 & 2.12&& \textbf{1.10} & \textbf{1.10} & 1.40&& \textbf{1.07} & \textbf{1.07} & 1.21&& \textbf{1.07} & \textbf{1.07} & 1.09\\
\texttt{b-100-to} & 100\%&& \textbf{1.16} & 1.24 & 2.70&& \textbf{1.15} & 1.16 & 1.45&& \textbf{1.12} & \textbf{1.12} & 1.31&& \textbf{1.10} & \textbf{1.10} & 1.11\\
\texttt{b-100-cl} & 100\%&& \textbf{1.17} & 1.18 & 2.39&& \textbf{1.14} & \textbf{1.14} & 1.49&& \textbf{1.12} & \textbf{1.12} & 1.20&& 1.12 & 1.10 & \textbf{1.07}\\
\texttt{b-100-ba} & 100\%&& 1.21 & \textbf{1.18} & 3.04&& 1.14 & \textbf{1.11} & 1.38&& \textbf{1.11} & \textbf{1.11} & \textbf{1.11}&& 1.11 & 1.11 & \textbf{1.03}\\
\midrule
\texttt{gka-er} & 99\%&& 1.25 & \textbf{1.24} & 2.59&& \textbf{1.20} & 1.24 & 1.44&& 1.20 & 1.20 & \textbf{1.19}&& 1.16 & 1.16 & \textbf{1.10}\\
\texttt{gka-to} & 100\%&& \textbf{1.29} & 1.32 & 2.61&& 1.25 & \textbf{1.18} & 1.62&& \textbf{1.18} & \textbf{1.18} & 1.26&& 1.12 & 1.15 & \textbf{1.08}\\
\texttt{gka-cl} & 100\%&& \textbf{1.22} & 1.24 & 3.14&& 1.18 & \textbf{1.15} & 1.39&& \textbf{1.12} & 1.17 & 1.24&& 1.09 & 1.12 & \textbf{1.07}\\
\texttt{gka-ba} & 100\%&& \textbf{1.27} & 1.28 & 3.19&& \textbf{1.20} & 1.24 & 2.25&& 1.20 & 1.16 & \textbf{1.12}&& 1.11 & 1.14 & \textbf{1.06}\\
\bottomrule
\end{tabular}

\caption{Observed approximation factors for anytime algorithms and heuristics at various timeouts.
For each dataset, the worst-case approximation ratio over its instances is reported.
Approximation ratios are with respect to \myoct on the reduced graph.
A checkmark denotes that exact solutions are found on all instances, if a dataset has no checkmark then the best approximation algorithm is bolded.
If \myoct could not be found within 10 minutes then the instance is not included; the percent of included data points is denoted in the \emph{represented} column.}
\label{table:heuristic-approx-synthetic}
\end{table*}

\subsection{Exact Solutions}
\begin{figure*}[!t]
\centering
\includegraphics[width=\linewidth]{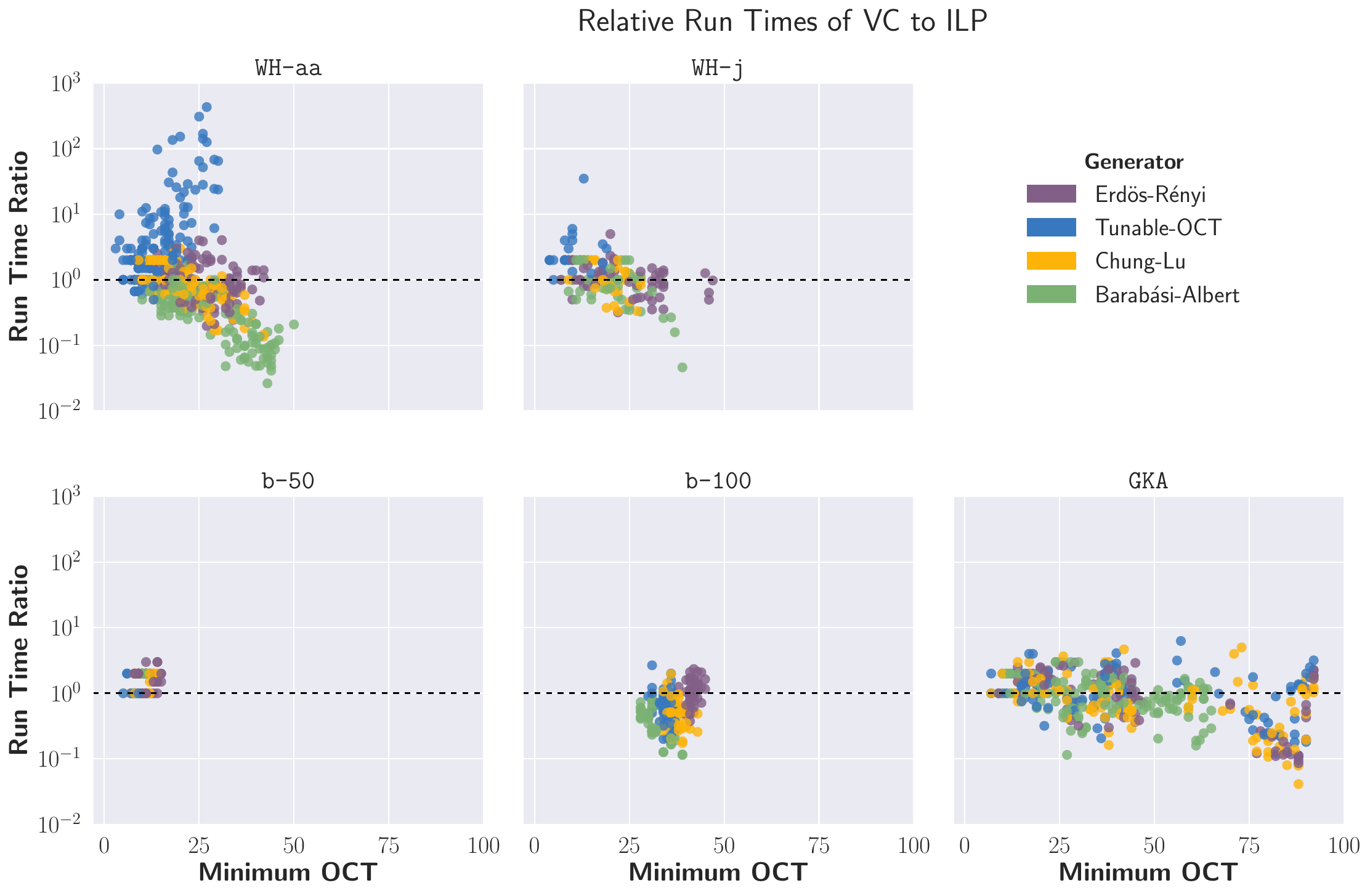}
\caption{Relative run times computed by dividing \textsf{VC} run times by those from \textsf{ILP} with one thread; the dashed line indicates equality. Data points reported over a synthetic corpus with five random generator seeds.}
\label{figure:exact_synthetic_comparison}
\end{figure*}

Expanding on Table~\ref{table:exact_summary}, we find that the choice between \ilp and \vc is heavily dataset- and generator-dependent.
Notably, in the first facet of Figure~\ref{figure:exact_synthetic_comparison} we find that Tunable-OCT data is difficult for the \vc-solver and CPLEX can solve the instances up to $1000\times$ faster.
However, the Barab\'asi-Albert instances are similarly easier for a \vc-solver.
The effectiveness of branch-and-reduce algorithms on instances with heavy-tailed degree distributions is well-known in folklore -- the few high-degree vertices in this network result in few branches, and the remaining (sparse) graph can sometimes be solved by reduction routines alone.
The results for \texttt{WH-j} and \texttt{GKA} are also split fairly evenly, although the latter has no clear split by generator.
Notably, \texttt{GKA} has instances with very large \myoct that are best solved with \ilp; in the rest of the data \vc tends to perform best as minimum OCT increases.
The \texttt{b-50} and \texttt{b-100} corpuses are best solved by \ilp and \vc, respectively.

Overall, \ilp generally does best on graphs with small minimum OCT ($\leq 25$), otherwise \vc returns faster solutions.
The results in \texttt{WH-aa} and \texttt{WH-j} also suggest that Tunable-OCT is best solved with \ilp; extended work could evaluate whether these datasets are particularly easy for \ilp or particularly hard for \vc.

\section{Conclusion}
\label{section:conclusion}
We experimentally evaluate state-of-the-art approaches to computing \myoctfull on the canonical \wh dataset, a new benchmark dataset from quantum annealing, and synthetic data generated using four random graph models which emphasize different structural properties.
Whereas previous experimental evaluations were limited to 61 graphs, each solved by modern methods within 1s, we utilize 116 existing benchmark instances, 840 frustrated cluster loop instances, and nearly 7000 synthetic instances from four different random graph generators.

On this significantly expanded corpus, we found that no single implementation dominates all scenarios.
Under the most extreme time constraint of 0.01s, sampling from simple heuristics provides the best solutions.
When using 0.1s, 1s, and 10s timeouts we find that a heuristic solution improved with structured work towards an exact solution performs best, including techniques such as iterative compression.
After 10s, though, the preprocessing overhead of using CPLEX pays off and an \ilp-solver dominates.

For significantly longer runs (a 10m timeout), we find \textsf{IC} uncompetitive and the choice between \vc- and \ilp-based solvers depends on the instance structure. We note that solvers may have inherent performance differences due to implementation language.
However, we expect that the effectiveness of solvers heavily depends on potency of their preprocessing routines -- CPLEX applies algebra-based reductions on the ILP before performing branch-and-cut, and Akiba and Iwata's \vc-solver applies reductions after each branch in their branch-and-reduce model.
More work is needed in identifying (in)effective reduction routines based on instance structure.
Results from the 2019 Parameterized Algorithms and Computational Experiments (PACE) challenge\footnote{\url{https://pacechallenge.org/2019/}} for \vcfull may be of particular interest towards this goal.

In addition to further study in preprocessing and structure of difficult instances, more flexible implementations are needed on the combinatorial algorithm front.
Algorithms such as \vc branch-and-reduce can be CPU-parallelized on a shared-memory system to potentially obtain super-linear speed-ups (witnessed by CPLEX in Appendix \ref{appendix:extended_results}).
Techniques such as iterative compression may also lend themselves to GPU-parallelization by executing an exponential number of disjoint \textsc{Minimum Cut} subroutines in parallel.

In addition to parallelism, implementations should also be provided as anytime algorithms whenever possible.
Many techniques create iteratively better lower bounds which can be completed into upper bounds with a simple rounding strategy.
These ``good enough'' solutions are directly useful for applications such as quantum computing, and the lower bounds may be useful in their own right.

\paragraph*{Acknowledgments}
\label{section:acknowledgments}
{
\small
This work supported in part by the Gordon $\&$ Betty Moore Foundations
Data-Driven Discovery Initiative through Grant GBMF4560 to Blair D. Sullivan,
and with Government support under and awarded by DoD, Air Force Office of
Scientific Research, National Defense Science and Engineering Graduate (NDSEG)
Fellowship, 32 CFR 168a to Timothy D. Goodrich.

}
\clearpage
\bibliography{Paper}

\clearpage
\appendix
\section{Implementation Details}
\label{appendix:implementation_details}
\subsection{Data Ingestion and Sanitization}
\label{subsection:data_implementation}

Original data comes from two sources. \texttt{Wernicke-\huffner} data is provided in the \huffner code download \cite{huffner-ic}, and \texttt{Beasley} and \texttt{GKA} data comes from Beasley's repository \cite{beasley}.

When parsing the graphs with Python we read them into a \netx graph and remove edges with weight zero (used to denote non-edges in some problems) and self-loops.
We then relabel the vertices to $\{0, \dots, n-1\}$.
To remove possible non-determinism in how vertices are relabeled, we specify that node labels are relabeled by lexicographical order of the original vertex labels, guaranteeing that each graph is always converted in the same way.

See the \textbf{Data} section of \replicability in our repository for information on how to use our scripts for automating this download and parsing process.

\subsection{Reduction Routines}
\label{subsection:preprocessing_implementation}
Reduction routines for preprocessing come from two papers: Wernicke \cite{wernicke2003algorithmic} and Akiba and Iwata \cite{akiba2016branch}.

While Wernicke originally implemented his reductions in Java, the code does not appear to have been open sourced.
We implement his reduction routines in Python3 with \netx.
Some care must be taken that these reductions operate deterministically so the results can be reproduced.
Specifically, reduction rules 4 and 6 require vertex cuts, which are returned in arbitrary order when computed by \netx; we convert the cuts to tuples and sort them by vertex label.
Additionally, reduction rules 7, 8, and 9 find and remove particular configurations in the graph based on degree 2 and 3 vertices; we sort these sets of vertices and the related neighborhoods.

For Akiba and Iwata's reduction routines, we modify their GitHub code \cite{ai-code} so that no branching is done after the first iteration of reduction routines, and the preprocessed graph is output instead.
To preprocess a graph, we apply Wernicke reductions first, then Akiba-Iwata reductions, and repeat until the graph does not change.
This was done primarily because some of Akiba-Iwata's reductions will not apply after the Wernicke reductions, simplifying the conversion from \vc to \myoct.

In order to make our experiments replicable, we verified that these reductions are determinstic by performing multiple rounds of preprocessing on different machines and checking that the resulting graphs were isomorphic, if small enough to be feasible, and had matching degree, triangle, and number of cliques sequences using the \netx \texttt{could\_be\_isomorphic} method otherwise.
To verify that these reductions are safe, we saved and verified a certificate (the odd cycle transversal) from each run of a solver that returned a feasible solution.

See the \textbf{Reductions} section of \replicability in our repository for information on how to run our scripts for applying these reduction routines.

\subsection{Heuristics}
\label{subsection:heuristic_implementation}
We implemented the heuristic ensemble in Modern C++14.
Given a graph file and a timeout, the ensemble will run greedy independent set (\mindeg), Luby's Algorithm (\luby), DFS 2-coloring (\dfs), and BFS 2-coloring (\bfs) in a round-robin fashion until the time limit is reached, returning the single best solution found by any heuristic.
See \cite{goodrich2017optimizing} for more on \mindeg, \cite{luby1986simple} for more on \luby, and \cite{wernicke2003algorithmic} for more on \dfs.

See the \textbf{Utilities/Heuristics Solver} section of \readme in our repository for information on how to run the heuristic ensemble solver.

\subsection{\huffner Improvements}
\label{appendix:huffner-improvements}
We implemented our improvements to \huffner's implementation \cite{huffner-ic} in Modern C++14, and rewrote the original solver to compile in C11.
By default, the \texttt{enum2col} solver is run, with the preprocessing level $p$ specified by the user: The default algorithm ($p=0$), the default algorithm with a heuristic bipartite subgraph starting the ordering ($p=1$), and the default algorithm with a heuristic bipartite subgraph starting the ordering and the remaining vertices sorted such edges are introduced as quickly as possible ($p=2$).

See the \textbf{Utilities/Iterative Compression Solver} section of \readme in our repository for information on how to download, install, and run this improved iterative compression solver.

\section{Frustrated Cluster Loop (FCL) Data}
\label{appendix:datasets}
\begin{table}[h!]
    \caption{
        Base clique sizes used to generate frustrated cluster loop (FCL) graphs appropriate for four types of quantum
        hardware. For each hardware type, the size of the base clique is between the size of the hardware's minimum
        clique-embedding and minimum biclique-embedding (inclusive).
    }
    \label{table:fcl-parameters}
    \begin{tabular}{lrrl}
        \toprule
        Hardware     & Min Clique-Embedding & Min Biclique-Embedding & FCL Base Clique Sizes                       \\
        \midrule
        D-Wave 2000Q &                   64 &                    128 &                        64, 80, 96, 112, 128 \\
        D-Wave 8000Q &                  128 &                    256 & 128, 144, 160, 176, 192, 208, 224, 240, 256 \\
        Pegasus-6    &                   62 &                    104 &                 62, 69, 76, 83, 90, 97, 104 \\
        Pegasus-12   &                  134 &                    248 &           134, 153, 172, 191, 210, 229, 248 \\
        \bottomrule
    \end{tabular}
\end{table}

\begin{figure}[h!]
    \includegraphics[width=0.8\textwidth]{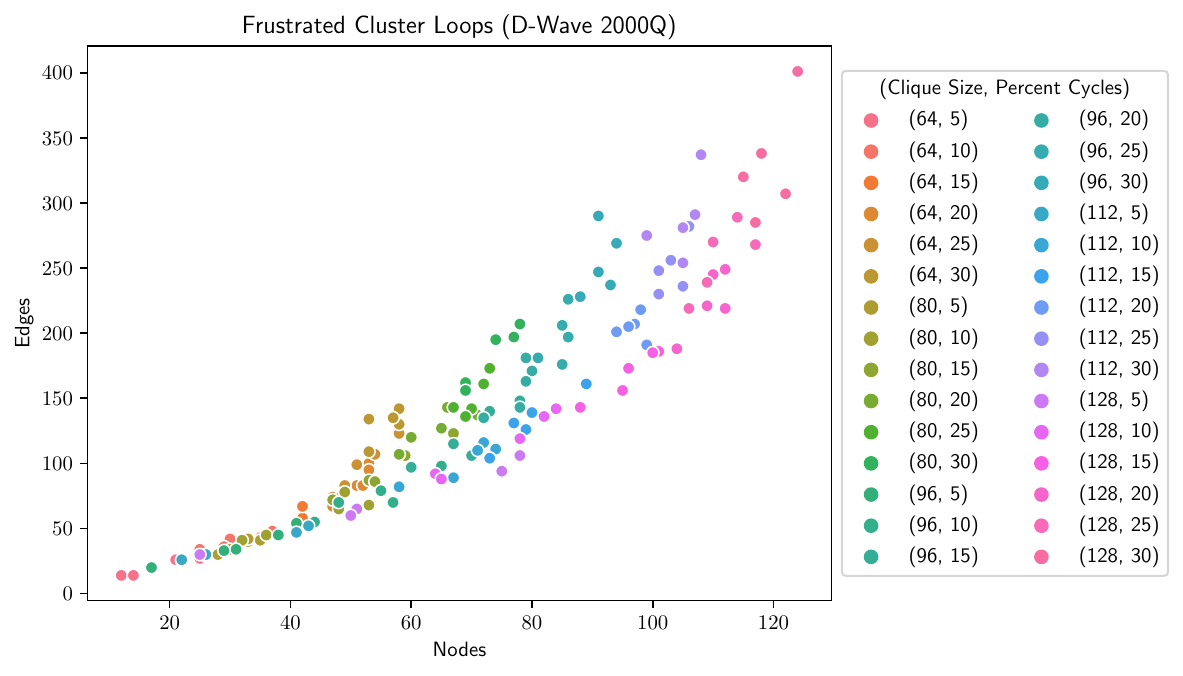}
    \caption{
        Instance sizes of frustrated cluster loop (FCL) graphs generated to be appropriate for the D-Wave 2000Q hardware.
        This hardware can embed a 64-clique and a 128-biclique, so we generated the FCLs from underlying cliques of size
        $64 \leq n \leq 128$ and varied the number of cycles from 5-30\% of the size of the clique. This ensures a corpus of
        FCLs with size $n \leq 128$, which is the problem space where \myoct-embedding methods would be used.
    }
    \label{figure:fcl}
\end{figure}

\section{Extended Results}
\label{appendix:extended_results}
\begin{figure*}[!t]
\includegraphics[height=0.2\textheight]{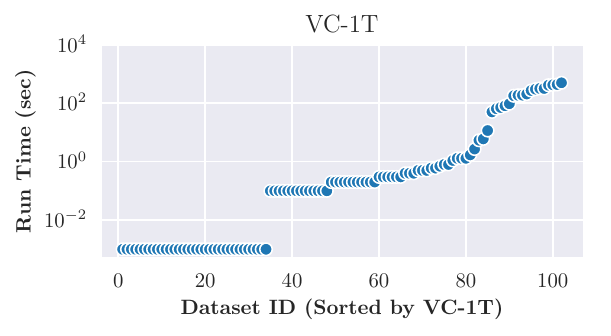}
\caption{Run times (log scale) of all quantum datasets when sorted in order of fastest to slowest when solved with the $\myoct \to \vc \to \ilp$ formulation and one thread (VC-1T).}
\label{figure:ilp_plot_vc_runtime}
\end{figure*}

\begin{figure*}[!b]
\includegraphics[height=0.2\textheight]{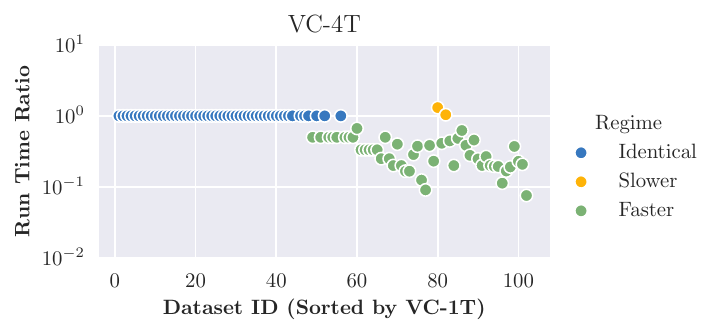}
\caption{The run time ratio (log scale) of a 4-threaded solver vs. a single-threaded solver using the $\myoct \to \vc \to \ilp$ formulation. Easier instances have identical solve times, and all but two of the harder instances benefit from the thread increase.}
\label{figure:ilp_plot_vc}
\end{figure*}

\subsection{ILP Solver Comparison}
\label{section:extended_ilp}

As noted in the introduction, previous work \cite{huffner2009algorithm, akiba2016branch} has conflicting reports on the effectiveness of \ilpfull (\ilp) solvers.
In this section we identify the best configuration for solving \myoct with \ilp by using \textsf{CPLEX} with either a formulation from $\myoct \to \ilp$ or $\myoct \to \vc \to \ilp$, and evaluate the impact of one thread vs. four.
It is well known that \textsf{GLPK} is not competitive with \textsf{CPLEX}, and the 4MB RAM limitations mentioned in \cite{huffner2009algorithm} are irrelevant given modern resources, therefore we do not consider these factors.

\begin{figure*}[!t]
\includegraphics[height=0.2\textheight]{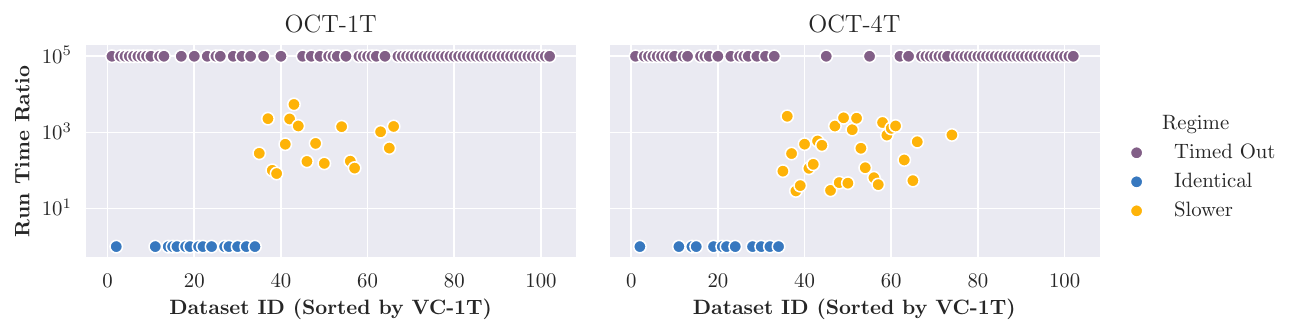}
\caption{The run time ratio (log scale) of the single- and four-threaded solvers using the $\myoct \to \ilp$ formulation compared to the single-threaded $\myoct \to \vc \to \ilp$ formulation (VC-1T). The majority of instances time out at 10 minutes, and no instance is solved faster than with VC-1T.}
\label{figure:ilp_plot_oct}
\end{figure*}

We find that using the $\myoct \to \vc \to \ilp$ formulation alone beats the alternative formulation, and that multithreading with this formulation can result in superlinear speedups.
Figure~\ref{figure:ilp_plot_vc_runtime} visualizes run times for all quantum datasets when sorted in order of fastest to solve with our canonical configuration, the $\myoct \to \vc \to \ilp$ formulation with one thread.
We find that in general (Figure~\ref{figure:ilp_plot_vc}), increasing to four threads is worthwhile and can result in a superlinear speedup; we observed speedups over $10\times$ for only a $4\times$ increase in cores.
Finally, we observe that the direct $\myoct \to \ilp$ formulation severely impacts \textsf{CPLEX} for the worse, regardless of threads used (Figure~\ref{figure:ilp_plot_oct}).

\subsection{Iterative Compression Heuristic Evaluation}
\label{section:extended_ic}
We implement our heuristic improvements in a version of \huffner's code improved for simplicity and compiler compatibility.
Recall the preprocessing parameter $p$ defined in Appendix~\ref{appendix:huffner-improvements}.
To evaluate these preprocessing options, we selected a subset of the data with a mixture of easy and difficult problems.
We then ran all three levels using 50 random seeds on each instance for each of four timeouts $\{0.01, 0.1, 1, 10\}$,
and report mean, standard deviation, and quintiles of the \myoct sizes found. Table~\ref{table:ic_full} depicts these results.
We find that $p=1$ always dominates $p=0$, especially in max and standard deviation, suggesting that certain orderings (avoided with $p=1$) are significantly disadvantageous for iterative compression.
Second, we find that $p=2$ tends to help on larger timeouts, where the run time cost of computing this ordering disappears.
Both observations can be seen in Table~\ref{table:ic_full}, where \texttt{aa-41} and \texttt{aa-42} benefit slightly from $p=2$, but \texttt{gka-2} and \texttt{gka-3} do not.

\begin{table*}[!h]
\center
\footnotesize
\setlength{\tabcolsep}{2pt}
\begin{tabular}{llrrrrrrrr}
\toprule
& & \multicolumn{7}{c}{\textbf{OCT Size}} \\
\cmidrule(lr){3-9}
\textbf{Dataset} & $p$ & \textbf{Mean} & \textbf{S.D.} & \textbf{Min} & \textbf{25\%} & \textbf{50\%} & \textbf{75\%} & \textbf{Max}\\
\midrule
\multicolumn{4}{l}{\textbf{Timeout: 0.01 (s)}}\\
\midrule
\texttt{aa-43}
& 0 &  21.4 &  3.0 &  18 &  19 &  21 &  23 &  29 \\
& 1 &  18.9 &  0.7 &  18 &  18 &  19 &  19 &  20 \\
& 2 &  19.5 &  0.6 &  18 &  19 &  20 &  20 &  20 \\
\texttt{aa-45}
& 0 &  29.6 &  4.2 &  22 &  27 &  30 &  32 &  41 \\
& 1 &  21.8 &  0.9 &  20 &  21 &  22 &  22 &  24 \\
& 2 &  22.7 &  0.7 &  21 &  22 &  23 &  23 &  24 \\
\texttt{gka-1}
& 0 &   9.0 &  0.0 &   9 &   9 &   9 &   9 &   9 \\
& 1 &   9.0 &  0.0 &   9 &   9 &   9 &   9 &   9 \\
& 2 &   9.0 &  0.0 &   9 &   9 &   9 &   9 &   9 \\
\midrule
\multicolumn{4}{l}{\textbf{Timeout: 0.1 (s)}}\\
\midrule
\texttt{aa-29}
& 0 &  35.9 &  16.2 &  21 &  23 &  32 &  39 &  97 \\
& 1 &  22.2 &   1.3 &  21 &  21 &  22 &  23 &  26 \\
& 2 &  23.9 &   1.7 &  21 &  23 &  24 &  25 &  28 \\
\midrule
\multicolumn{4}{l}{\textbf{Timeout: 1 (s)}}\\
\midrule
\texttt{aa-42}
& 0 &  51.8 &  13.4 &  33 &  42 &  49 &  59 &  85 \\
& 1 &  31.0 &   1.2 &  30 &  30 &  31 &  32 &  35 \\
& 2 &  31.9 &   0.3 &  31 &  32 &  32 &  32 &  32 \\
\texttt{gka-2}
& 0 &  16.8 &   0.9 &  16 &  16 &  17 &  17 &  19 \\
& 1 &  16.0 &   0.0 &  16 &  16 &  16 &  16 &  16 \\
& 2 &  16.0 &   0.0 &  16 &  16 &  16 &  16 &  16 \\
\midrule
\multicolumn{4}{l}{\textbf{Timeout: 10 (s)}}\\
\midrule
\texttt{aa-32}
& 0 &  31.6 &   2.4 &  30 &  30 &  30 &  32 &   38 \\
& 1 &  30.8 &   0.9 &  30 &  30 &  31 &  31 &   33 \\
& 2 &  30.8 &   0.4 &  30 &  31 &  31 &  31 &   31 \\
\texttt{aa-41}
& 0 &  67.0 &  14.4 &  48 &  55 &  64 &  75 &  110 \\
& 1 &  41.4 &   1.2 &  40 &  40 &  41 &  42 &   44 \\
& 2 &  41.2 &   0.7 &  40 &  41 &  41 &  41 &   43 \\
\texttt{gka-24}
& 0 &  47.1 &   1.6 &  44 &  46 &  47 &  48 &   50 \\
& 1 &  37.9 &   0.3 &  37 &  38 &  38 &  38 &   38 \\
& 2 &  37.9 &   0.3 &  37 &  38 &  38 &  38 &   38 \\
\texttt{gka-25}
& 0 &  56.2 &   2.6 &  49 &  55 &  56 &  58 &   62 \\
& 1 &  44.0 &   0.0 &  44 &  44 &  44 &  44 &   44 \\
& 2 &  44.0 &   0.0 &  44 &  44 &  44 &  44 &   44 \\
\texttt{gka-26}
& 0 &  57.0 &   2.3 &  52 &  55 &  57 &  58 &   62 \\
& 1 &  43.8 &   0.4 &  43 &  44 &  44 &  44 &   44 \\
& 2 &  43.8 &   0.4 &  43 &  44 &  44 &  44 &   44 \\
\texttt{gka-27}
& 0 &  72.3 &   1.7 &  68 &  71 &  72 &  74 &   75 \\
& 1 &  63.0 &   0.1 &  62 &  63 &  63 &  63 &   63 \\
& 2 &  63.0 &   0.0 &  63 &  63 &  63 &  63 &   63 \\
\texttt{gka-28}
& 0 &  78.0 &   1.2 &  75 &  77 &  78 &  79 &   80 \\
& 1 &  72.0 &   0.2 &  71 &  72 &  72 &  72 &   73 \\
& 2 &  72.0 &   0.2 &  72 &  72 &  72 &  72 &   73 \\
\texttt{gka-29}
& 0 &  82.4 &   0.8 &  81 &  82 &  82 &  83 &   84 \\
& 1 &  77.0 &   0.0 &  77 &  77 &  77 &  77 &   77 \\
& 2 &  77.0 &   0.0 &  77 &  77 &  77 &  77 &   77 \\
\texttt{gka-3}
& 0 &  25.2 &   1.6 &  23 &  24 &  25 &  26 &   29 \\
& 1 &  23.3 &   0.5 &  23 &  23 &  23 &  24 &   24 \\
& 2 &  23.4 &   0.5 &  23 &  23 &  23 &  24 &   24 \\
\bottomrule
\end{tabular}

\caption{Heuristic solution sizes for select datasets at the three levels of preprocessing ($p \in \{0, 1, 2\}$) for \textsf{Improved \huffner}. The timeout level on each dataset was taken to be the maximum time in $\{0.01, 0.1, 1, 10\}$ less than the time \huffner at $p = 0$ took to find an exact solution. Mean, standard deviation, and quintiles are computed over 50 samples per timeout, dataset, and preprocessing level.}
\label{table:ic_full}
\end{table*}

\subsection{Additional Tables}
\begin{table*}[!h]
\tabcolsep = 1.2mm
\center
\scriptsize
\begin{tabular}{lrcrrrcrrrcrrrcrrrl}
\toprule
& \multicolumn{2}{c}{\textbf{Timeout:}} &
\multicolumn{3}{c}{\textbf{0.01(s)}} & \phantom{Q} &
\multicolumn{3}{c}{\textbf{0.1(s)}} & \phantom{Q} &
\multicolumn{3}{c}{\textbf{1(s)}} & \phantom{Q} &
\multicolumn{3}{c}{\textbf{10(s)}}\\
\cmidrule(lr){4-6}
\cmidrule(lr){8-10}
\cmidrule(lr){12-14}
\cmidrule(lr){16-18}
\textbf{Dataset} & $OPT'$ & &
\textsf{HE} & \textsf{IC} & \textsf{ILP} &  &
\textsf{HE} & \textsf{IC} & \textsf{ILP} &  &
\textsf{HE} & \textsf{IC} & \textsf{ILP} &  &
\textsf{HE} & \textsf{IC} & \textsf{ILP}\\
\midrule\multicolumn{10}{l}{\textbf{Wernicke-\huffner Afro-American Graphs}}\\
\midrule\texttt{aa-10} & 6 & &
7 & \textBF{6} & 8 & &
6 & \textBF{6} & \textBF{6} & &
6 & \textBF{6} & \textBF{6} & &
6 & \textBF{6} & \textBF{6}\\

\texttt{aa-11} & 11 & &
12 & \textBF{11} & 15 & &
12 & \textBF{11} & 13 & &
11 & \textBF{11} & \textBF{11} & &
11 & \textBF{11} & \textBF{11}\\

\texttt{aa-13} & 12 & &
15 & \textBF{12} & 22 & &
15 & \textBF{12} & 12 & &
14 & \textBF{12} & \textBF{12} & &
13 & \textBF{12} & \textBF{12}\\

\texttt{aa-14} & 19 & &
21 & 19 & 35 & &
19 & 19 & 27 & &
19 & 19 & \textBF{19} & &
19 & \textBF{19} & \textBF{19}\\

\texttt{aa-15} & 6 & &
8 & \textBF{6} & 12 & &
6 & \textBF{6} & \textBF{6} & &
6 & \textBF{6} & \textBF{6} & &
6 & \textBF{6} & \textBF{6}\\

\texttt{aa-16} & 0 & &
0 & \textBF{0} & \textBF{0} & &
0 & \textBF{0} & \textBF{0} & &
0 & \textBF{0} & \textBF{0} & &
0 & \textBF{0} & \textBF{0}\\

\texttt{aa-17} & 25 & &
28 & 28 & 41 & &
28 & 27 & 34 & &
26 & 27 & \textBF{25} & &
26 & \textBF{25} & \textBF{25}\\

\texttt{aa-18} & 14 & &
18 & \textBF{14} & 21 & &
17 & \textBF{14} & 14 & &
14 & \textBF{14} & \textBF{14} & &
14 & \textBF{14} & \textBF{14}\\

\texttt{aa-19} & 19 & &
22 & 22 & 38 & &
21 & 21 & 20 & &
21 & 19 & \textBF{19} & &
21 & \textBF{19} & \textBF{19}\\

\texttt{aa-20} & 19 & &
23 & 24 & 129 & &
22 & 20 & 25 & &
22 & 20 & \textBF{19} & &
22 & 20 & \textBF{19}\\

\texttt{aa-21} & 0 & &
0 & \textBF{0} & \textBF{0} & &
0 & \textBF{0} & \textBF{0} & &
0 & \textBF{0} & \textBF{0} & &
0 & \textBF{0} & \textBF{0}\\

\texttt{aa-22} & 16 & &
17 & 17 & 37 & &
17 & 16 & 16 & &
17 & \textBF{16} & \textBF{16} & &
17 & \textBF{16} & \textBF{16}\\

\texttt{aa-23} & 18 & &
21 & 21 & 35 & &
21 & \textBF{18} & 19 & &
19 & \textBF{18} & \textBF{18} & &
19 & \textBF{18} & \textBF{18}\\

\texttt{aa-24} & 21 & &
25 & 27 & 150 & &
25 & 24 & 46 & &
24 & \textBF{21} & \textBF{21} & &
24 & \textBF{21} & \textBF{21}\\

\texttt{aa-25} & 0 & &
0 & \textBF{0} & \textBF{0} & &
0 & \textBF{0} & \textBF{0} & &
0 & \textBF{0} & \textBF{0} & &
0 & \textBF{0} & \textBF{0}\\

\texttt{aa-26} & 12 & &
13 & \textBF{12} & 14 & &
12 & \textBF{12} & \textBF{12} & &
12 & \textBF{12} & \textBF{12} & &
12 & \textBF{12} & \textBF{12}\\

\texttt{aa-27} & 11 & &
14 & \textBF{11} & 13 & &
14 & \textBF{11} & 11 & &
13 & \textBF{11} & \textBF{11} & &
12 & \textBF{11} & \textBF{11}\\

\texttt{aa-28} & 27 & &
33 & 32 & 106 & &
31 & 29 & 32 & &
30 & \textBF{27} & \textBF{27} & &
30 & 28 & \textBF{27}\\

\texttt{aa-29} & 21 & &
26 & 26 & 169 & &
22 & 24 & 23 & &
22 & \textBF{21} & \textBF{21} & &
22 & \textBF{21} & \textBF{21}\\

\texttt{aa-30} & 4 & &
4 & \textBF{4} & \textBF{4} & &
4 & \textBF{4} & \textBF{4} & &
4 & \textBF{4} & \textBF{4} & &
4 & \textBF{4} & \textBF{4}\\

\texttt{aa-31} & 0 & &
0 & \textBF{0} & \textBF{0} & &
0 & \textBF{0} & \textBF{0} & &
0 & \textBF{0} & \textBF{0} & &
0 & \textBF{0} & \textBF{0}\\

\texttt{aa-32} & 30 & &
39 & 35 & 47 & &
32 & 35 & 37 & &
32 & 33 & \textBF{30} & &
32 & 31 & \textBF{30}\\

\texttt{aa-33} & 4 & &
4 & \textBF{4} & 31 & &
4 & \textBF{4} & \textBF{4} & &
4 & \textBF{4} & \textBF{4} & &
4 & \textBF{4} & \textBF{4}\\

\texttt{aa-34} & 13 & &
16 & \textBF{13} & 27 & &
13 & \textBF{13} & 13 & &
13 & \textBF{13} & \textBF{13} & &
13 & \textBF{13} & \textBF{13}\\

\texttt{aa-35} & 10 & &
10 & \textBF{10} & \textBF{10} & &
10 & \textBF{10} & \textBF{10} & &
10 & \textBF{10} & \textBF{10} & &
10 & \textBF{10} & \textBF{10}\\

\texttt{aa-36} & 7 & &
7 & \textBF{7} & 18 & &
7 & \textBF{7} & \textBF{7} & &
7 & \textBF{7} & \textBF{7} & &
7 & \textBF{7} & \textBF{7}\\

\texttt{aa-37} & 4 & &
5 & \textBF{4} & 7 & &
4 & \textBF{4} & \textBF{4} & &
4 & \textBF{4} & \textBF{4} & &
4 & \textBF{4} & \textBF{4}\\

\texttt{aa-38} & 26 & &
33 & 31 & 102 & &
31 & 27 & 35 & &
31 & \textBF{26} & \textBF{26} & &
31 & \textBF{26} & \textBF{26}\\

\texttt{aa-39} & 23 & &
28 & 27 & 40 & &
26 & 26 & 26 & &
24 & 25 & \textBF{23} & &
24 & \textBF{23} & \textBF{23}\\

\texttt{aa-40} & 22 & &
25 & 28 & 33 & &
22 & 23 & 28 & &
22 & \textBF{22} & \textBF{22} & &
22 & \textBF{22} & \textBF{22}\\

\texttt{aa-41} & 40 & &
50 & 46 & 172 & &
48 & 45 & 60 & &
46 & 45 & \textBF{40} & &
46 & 41 & \textBF{40}\\

\texttt{aa-42} & 30 & &
36 & 35 & 160 & &
36 & 33 & 43 & &
34 & 32 & \textBF{30} & &
34 & \textBF{30} & \textBF{30}\\

\texttt{aa-43} & 18 & &
19 & 19 & 20 & &
19 & 19 & 18 & &
19 & 19 & \textBF{18} & &
19 & 19 & \textBF{18}\\

\texttt{aa-44} & 10 & &
10 & \textBF{10} & 15 & &
10 & \textBF{10} & 12 & &
10 & \textBF{10} & \textBF{10} & &
10 & \textBF{10} & \textBF{10}\\

\texttt{aa-45} & 20 & &
22 & 22 & \textBF{20} & &
21 & 22 & 20 & &
21 & \textBF{20} & \textBF{20} & &
21 & \textBF{20} & \textBF{20}\\

\texttt{aa-46} & 13 & &
18 & 13 & 37 & &
16 & \textBF{13} & \textBF{13} & &
14 & \textBF{13} & \textBF{13} & &
14 & \textBF{13} & \textBF{13}\\

\texttt{aa-47} & 13 & &
14 & \textBF{13} & 15 & &
14 & \textBF{13} & 13 & &
13 & \textBF{13} & \textBF{13} & &
13 & \textBF{13} & \textBF{13}\\

\texttt{aa-48} & 17 & &
19 & 18 & 21 & &
17 & 17 & 17 & &
17 & \textBF{17} & \textBF{17} & &
17 & \textBF{17} & \textBF{17}\\

\texttt{aa-49} & 0 & &
0 & \textBF{0} & \textBF{0} & &
0 & \textBF{0} & \textBF{0} & &
0 & \textBF{0} & \textBF{0} & &
0 & \textBF{0} & \textBF{0}\\

\texttt{aa-50} & 18 & &
21 & 19 & 27 & &
21 & 18 & 19 & &
20 & \textBF{18} & \textBF{18} & &
19 & \textBF{18} & \textBF{18}\\

\texttt{aa-51} & 11 & &
13 & \textBF{11} & 16 & &
11 & \textBF{11} & \textBF{11} & &
11 & \textBF{11} & \textBF{11} & &
11 & \textBF{11} & \textBF{11}\\

\texttt{aa-52} & 12 & &
13 & \textBF{12} & 15 & &
12 & \textBF{12} & 12 & &
12 & \textBF{12} & \textBF{12} & &
12 & \textBF{12} & \textBF{12}\\

\texttt{aa-53} & 12 & &
15 & 12 & 18 & &
14 & \textBF{12} & 13 & &
14 & \textBF{12} & \textBF{12} & &
13 & \textBF{12} & \textBF{12}\\

\texttt{aa-54} & 12 & &
14 & 12 & 21 & &
14 & \textBF{12} & 14 & &
13 & \textBF{12} & \textBF{12} & &
12 & \textBF{12} & \textBF{12}\\

\midrule\multicolumn{10}{l}{\textbf{Wernicke-\huffner Japanese Graphs}}\\
\midrule\texttt{j-10} & 3 & &
3 & \textBF{3} & \textBF{3} & &
3 & \textBF{3} & \textBF{3} & &
3 & \textBF{3} & \textBF{3} & &
3 & \textBF{3} & \textBF{3}\\

\texttt{j-11} & 5 & &
5 & \textBF{5} & \textBF{5} & &
5 & \textBF{5} & \textBF{5} & &
5 & \textBF{5} & \textBF{5} & &
5 & \textBF{5} & \textBF{5}\\

\texttt{j-13} & 4 & &
4 & \textBF{4} & \textBF{4} & &
4 & \textBF{4} & \textBF{4} & &
4 & \textBF{4} & \textBF{4} & &
4 & \textBF{4} & \textBF{4}\\

\texttt{j-14} & 3 & &
3 & \textBF{3} & \textBF{3} & &
3 & \textBF{3} & \textBF{3} & &
3 & \textBF{3} & \textBF{3} & &
3 & \textBF{3} & \textBF{3}\\

\texttt{j-15} & 0 & &
0 & \textBF{0} & \textBF{0} & &
0 & \textBF{0} & \textBF{0} & &
0 & \textBF{0} & \textBF{0} & &
0 & \textBF{0} & \textBF{0}\\

\texttt{j-16} & 0 & &
0 & \textBF{0} & \textBF{0} & &
0 & \textBF{0} & \textBF{0} & &
0 & \textBF{0} & \textBF{0} & &
0 & \textBF{0} & \textBF{0}\\

\texttt{j-17} & 10 & &
10 & \textBF{10} & 16 & &
10 & \textBF{10} & \textBF{10} & &
10 & \textBF{10} & \textBF{10} & &
10 & \textBF{10} & \textBF{10}\\

\texttt{j-18} & 9 & &
10 & 10 & 12 & &
10 & 10 & \textBF{9} & &
9 & \textBF{9} & \textBF{9} & &
9 & \textBF{9} & \textBF{9}\\

\texttt{j-19} & 3 & &
3 & \textBF{3} & \textBF{3} & &
3 & \textBF{3} & \textBF{3} & &
3 & \textBF{3} & \textBF{3} & &
3 & \textBF{3} & \textBF{3}\\

\texttt{j-20} & 0 & &
0 & \textBF{0} & \textBF{0} & &
0 & \textBF{0} & \textBF{0} & &
0 & \textBF{0} & \textBF{0} & &
0 & \textBF{0} & \textBF{0}\\

\texttt{j-21} & 2 & &
2 & \textBF{2} & \textBF{2} & &
2 & \textBF{2} & \textBF{2} & &
2 & \textBF{2} & \textBF{2} & &
2 & \textBF{2} & \textBF{2}\\

\texttt{j-22} & 9 & &
9 & \textBF{9} & \textBF{9} & &
9 & \textBF{9} & \textBF{9} & &
9 & \textBF{9} & \textBF{9} & &
9 & \textBF{9} & \textBF{9}\\

\texttt{j-23} & 19 & &
19 & 19 & \textBF{19} & &
19 & 19 & 19 & &
19 & \textBF{19} & \textBF{19} & &
19 & \textBF{19} & \textBF{19}\\

\texttt{j-24} & 3 & &
3 & \textBF{3} & \textBF{3} & &
3 & \textBF{3} & \textBF{3} & &
3 & \textBF{3} & \textBF{3} & &
3 & \textBF{3} & \textBF{3}\\

\texttt{j-25} & 0 & &
0 & \textBF{0} & \textBF{0} & &
0 & \textBF{0} & \textBF{0} & &
0 & \textBF{0} & \textBF{0} & &
0 & \textBF{0} & \textBF{0}\\

\texttt{j-26} & 6 & &
6 & \textBF{6} & 7 & &
6 & \textBF{6} & \textBF{6} & &
6 & \textBF{6} & \textBF{6} & &
6 & \textBF{6} & \textBF{6}\\

\texttt{j-28} & 13 & &
13 & 16 & 16 & &
13 & 13 & \textBF{13} & &
13 & \textBF{13} & \textBF{13} & &
13 & \textBF{13} & \textBF{13}\\

\bottomrule
\end{tabular}

\caption{Heuristic solution sizes after 0.01, 0.1, 1, and 10 seconds for the heuristic ensemble (\textsf{HE}),  iterative compression \textsf{IC}, and \textsf{CPLEX} (\textsf{ILP}) when run on \texttt{WH} datasets. Highlighting indicates instances where the solver found an exact solution before the provided time limit.}
\label{table:heuristic1}
\end{table*}

\begin{table*}[!h]
\tabcolsep = 1.2mm
\center
\scriptsize
\begin{tabular}{lrcrrrcrrrcrrrcrrrl}
\toprule
& \multicolumn{2}{c}{\textbf{Timeout:}} &
\multicolumn{3}{c}{\textbf{0.01(s)}} & \phantom{Q} &
\multicolumn{3}{c}{\textbf{0.1(s)}} & \phantom{Q} &
\multicolumn{3}{c}{\textbf{1(s)}} & \phantom{Q} &
\multicolumn{3}{c}{\textbf{10(s)}}\\
\cmidrule(lr){4-6}
\cmidrule(lr){8-10}
\cmidrule(lr){12-14}
\cmidrule(lr){16-18}
\textbf{Dataset} & $OPT$ & &
\textsf{HE} & \textsf{IC} & \textsf{ILP} &  &
\textsf{HE} & \textsf{IC} & \textsf{ILP} &  &
\textsf{HE} & \textsf{IC} & \textsf{ILP} &  &
\textsf{HE} & \textsf{IC} & \textsf{ILP}\\
\midrule\multicolumn{10}{l}{\textbf{Beasley 50-Vertex Graphs}}\\
\midrule\texttt{b-50-1} & 11 & &
12 & 11 & 15 & &
11 & 11 & 12 & &
11 & \textBF{11} & \textBF{11} & &
11 & \textBF{11} & \textBF{11}\\

\texttt{b-50-2} & 11 & &
12 & 11 & 13 & &
12 & \textBF{11} & \textBF{11} & &
11 & \textBF{11} & \textBF{11} & &
11 & \textBF{11} & \textBF{11}\\

\texttt{b-50-3} & 14 & &
15 & 15 & 16 & &
14 & 14 & 14 & &
14 & \textBF{14} & \textBF{14} & &
14 & \textBF{14} & \textBF{14}\\

\texttt{b-50-4} & 11 & &
12 & 12 & 14 & &
11 & \textBF{11} & 11 & &
11 & \textBF{11} & \textBF{11} & &
11 & \textBF{11} & \textBF{11}\\

\texttt{b-50-5} & 13 & &
13 & 13 & 16 & &
13 & 13 & 15 & &
13 & \textBF{13} & \textBF{13} & &
13 & \textBF{13} & \textBF{13}\\

\texttt{b-50-6} & 9 & &
9 & \textBF{9} & 14 & &
9 & \textBF{9} & 9 & &
9 & \textBF{9} & \textBF{9} & &
9 & \textBF{9} & \textBF{9}\\

\texttt{b-50-7} & 9 & &
10 & \textBF{9} & 12 & &
9 & \textBF{9} & \textBF{9} & &
9 & \textBF{9} & \textBF{9} & &
9 & \textBF{9} & \textBF{9}\\

\texttt{b-50-8} & 14 & &
15 & 15 & 20 & &
15 & 15 & 17 & &
15 & \textBF{14} & \textBF{14} & &
14 & \textBF{14} & \textBF{14}\\

\texttt{b-50-9} & 12 & &
13 & 12 & 16 & &
13 & 12 & 12 & &
13 & \textBF{12} & \textBF{12} & &
12 & \textBF{12} & \textBF{12}\\

\texttt{b-50-10} & 11 & &
12 & 11 & 13 & &
12 & \textBF{11} & 11 & &
11 & \textBF{11} & \textBF{11} & &
11 & \textBF{11} & \textBF{11}\\

\midrule\multicolumn{10}{l}{\textbf{Beasley 100-Vertex Graphs}}\\
\midrule\texttt{b-100-1} & 41 & &
43 & 44 & 46 & &
43 & 44 & 46 & &
43 & 43 & 43 & &
43 & 43 & 42\\

\texttt{b-100-2} & 42 & &
44 & 44 & 47 & &
43 & 44 & 47 & &
43 & 43 & 43 & &
43 & 43 & 42\\

\texttt{b-100-3} & 42 & &
46 & 45 & 49 & &
45 & 45 & 49 & &
44 & 44 & 46 & &
44 & 44 & 42\\

\texttt{b-100-4} & 41 & &
43 & 45 & 45 & &
42 & 42 & 45 & &
42 & 42 & 42 & &
42 & 42 & 42\\

\texttt{b-100-5} & 42 & &
45 & 44 & 47 & &
44 & 44 & 47 & &
43 & 43 & 45 & &
42 & 43 & 43\\

\texttt{b-100-6} & 43 & &
45 & 46 & 48 & &
45 & 45 & 48 & &
45 & 45 & 45 & &
45 & 45 & 44\\

\texttt{b-100-7} & 42 & &
45 & 45 & 47 & &
44 & 44 & 47 & &
44 & 44 & 43 & &
44 & 44 & 42\\

\texttt{b-100-8} & 43 & &
47 & 46 & 48 & &
45 & 45 & 48 & &
45 & 45 & 47 & &
45 & 45 & 43\\

\texttt{b-100-9} & 44 & &
46 & 47 & 49 & &
45 & 45 & 49 & &
45 & 45 & 46 & &
45 & 45 & 44\\

\texttt{b-100-10} & 44 & &
46 & 48 & 50 & &
45 & 46 & 50 & &
45 & 45 & 46 & &
45 & 45 & 45\\

\midrule\multicolumn{10}{l}{\textbf{GKA Graphs}}\\
\midrule\texttt{gka-1} & 9 & &
9 & \textBF{9} & 12 & &
9 & \textBF{9} & \textBF{9} & &
9 & \textBF{9} & \textBF{9} & &
9 & \textBF{9} & \textBF{9}\\

\texttt{gka-2} & 16 & &
16 & 16 & 19 & &
16 & 16 & 17 & &
16 & 16 & \textBF{16} & &
16 & \textBF{16} & \textBF{16}\\

\texttt{gka-3} & 23 & &
24 & 24 & 24 & &
24 & 24 & 24 & &
24 & 23 & \textBF{23} & &
23 & 23 & \textBF{23}\\

\texttt{gka-4} & 28 & &
31 & 31 & 33 & &
31 & 30 & 33 & &
30 & 30 & 28 & &
30 & 30 & \textBF{28}\\

\texttt{gka-5} & 22 & &
22 & 22 & 23 & &
22 & 22 & 23 & &
22 & 22 & \textBF{22} & &
22 & \textBF{22} & \textBF{22}\\

\texttt{gka-6} & 16 & &
16 & 16 & \textBF{16} & &
16 & \textBF{16} & \textBF{16} & &
16 & \textBF{16} & \textBF{16} & &
16 & \textBF{16} & \textBF{16}\\

\texttt{gka-7} & 18 & &
18 & \textBF{18} & \textBF{18} & &
18 & \textBF{18} & \textBF{18} & &
18 & \textBF{18} & \textBF{18} & &
18 & \textBF{18} & \textBF{18}\\

\texttt{gka-8} & 28 & &
34 & 33 & 33 & &
33 & 33 & 33 & &
31 & 32 & 28 & &
31 & 32 & \textBF{28}\\

\texttt{gka-9} & 2 & &
2 & \textBF{2} & \textBF{2} & &
2 & \textBF{2} & \textBF{2} & &
2 & \textBF{2} & \textBF{2} & &
2 & \textBF{2} & \textBF{2}\\

\texttt{gka-10} & 8 & &
8 & \textBF{8} & \textBF{8} & &
8 & \textBF{8} & \textBF{8} & &
8 & \textBF{8} & \textBF{8} & &
8 & \textBF{8} & \textBF{8}\\

\texttt{gka-11} & 10 & &
10 & \textBF{10} & \textBF{10} & &
10 & \textBF{10} & \textBF{10} & &
10 & \textBF{10} & \textBF{10} & &
10 & \textBF{10} & \textBF{10}\\

\texttt{gka-12} & 20 & &
20 & \textBF{20} & \textBF{20} & &
20 & \textBF{20} & \textBF{20} & &
20 & \textBF{20} & \textBF{20} & &
20 & \textBF{20} & \textBF{20}\\

\texttt{gka-13} & 16 & &
16 & \textBF{16} & \textBF{16} & &
16 & \textBF{16} & \textBF{16} & &
16 & \textBF{16} & \textBF{16} & &
16 & \textBF{16} & \textBF{16}\\

\texttt{gka-14} & 22 & &
22 & \textBF{22} & \textBF{22} & &
22 & \textBF{22} & \textBF{22} & &
22 & \textBF{22} & \textBF{22} & &
22 & \textBF{22} & \textBF{22}\\

\texttt{gka-15} & 0 & &
0 & \textBF{0} & \textBF{0} & &
0 & \textBF{0} & \textBF{0} & &
0 & \textBF{0} & \textBF{0} & &
0 & \textBF{0} & \textBF{0}\\

\texttt{gka-16} & 0 & &
0 & \textBF{0} & \textBF{0} & &
0 & \textBF{0} & \textBF{0} & &
0 & \textBF{0} & \textBF{0} & &
0 & \textBF{0} & \textBF{0}\\

\texttt{gka-17} & 0 & &
0 & \textBF{0} & \textBF{0} & &
0 & \textBF{0} & \textBF{0} & &
0 & \textBF{0} & \textBF{0} & &
0 & \textBF{0} & \textBF{0}\\

\texttt{gka-18} & 2 & &
2 & \textBF{2} & \textBF{2} & &
2 & \textBF{2} & \textBF{2} & &
2 & \textBF{2} & \textBF{2} & &
2 & \textBF{2} & \textBF{2}\\

\texttt{gka-19} & 31 & &
31 & 31 & 32 & &
31 & \textBF{31} & \textBF{31} & &
31 & \textBF{31} & \textBF{31} & &
31 & \textBF{31} & \textBF{31}\\

\texttt{gka-20} & 39 & &
39 & 39 & 40 & &
39 & 39 & 39 & &
39 & \textBF{39} & \textBF{39} & &
39 & \textBF{39} & \textBF{39}\\

\texttt{gka-21} & 40 & &
41 & 42 & 41 & &
41 & 42 & 41 & &
41 & 41 & \textBF{40} & &
40 & 41 & \textBF{40}\\

\texttt{gka-22} & 43 & &
46 & 46 & 49 & &
44 & 46 & 46 & &
44 & 46 & 43 & &
43 & 45 & \textBF{43}\\

\texttt{gka-23} & 46 & &
48 & 48 & 52 & &
48 & 48 & 51 & &
47 & 47 & 47 & &
46 & 46 & 46\\

\texttt{gka-24} & 37 & &
39 & 38 & 44 & &
38 & 38 & 43 & &
38 & 38 & 39 & &
37 & 38 & 37\\

\texttt{gka-25} & 42 & &
45 & 46 & 47 & &
44 & 44 & 47 & &
44 & 44 & 45 & &
44 & 44 & 43\\

\texttt{gka-26} & 43 & &
44 & 44 & 50 & &
44 & 44 & 50 & &
44 & 44 & 43 & &
44 & 44 & 43\\

\texttt{gka-27} & 62 & &
65 & 64 & 75 & &
64 & 63 & 65 & &
63 & 63 & 63 & &
63 & 63 & 63\\

\texttt{gka-28} & 70 & &
73 & 73 & 94 & &
72 & 72 & 74 & &
72 & 72 & 71 & &
72 & 72 & 71\\

\texttt{gka-29} & 77 & &
77 & 77 & 95 & &
77 & 77 & 80 & &
77 & 77 & 79 & &
77 & 77 & 78\\

\texttt{gka-30} & 82 & &
83 & 84 & 96 & &
83 & 83 & 85 & &
83 & 83 & 83 & &
83 & 83 & 82\\

\texttt{gka-31} & 85 & &
86 & 87 & 95 & &
86 & 87 & 88 & &
86 & 86 & 87 & &
86 & 86 & 85\\

\texttt{gka-32} & 88 & &
89 & 89 & 97 & &
89 & 89 & 90 & &
89 & 89 & 89 & &
88 & 88 & 88\\

\texttt{gka-33} & 90 & &
91 & 91 & 98 & &
91 & 90 & 98 & &
90 & 90 & 90 & &
90 & \textBF{90} & \textBF{90}\\

\texttt{gka-34} & 92 & &
93 & 93 & 98 & &
93 & 92 & 98 & &
93 & \textBF{92} & 92 & &
92 & \textBF{92} & \textBF{92}\\

\texttt{gka-35} & 0 & &
0 & \textBF{0} & \textBF{0} & &
0 & \textBF{0} & \textBF{0} & &
0 & \textBF{0} & \textBF{0} & &
0 & \textBF{0} & \textBF{0}\\

\bottomrule
\end{tabular}

\caption{Heuristic solution sizes after 0.01, 0.1, 1, and 10 seconds for the heuristic ensemble (\textsf{HE}),  iterative compression \textsf{IC}, and \textsf{CPLEX} (\textsf{ILP}) when run on \texttt{Beasley} and \texttt{GKA} datasets. Highlighting indicates instances where the solver found an exact solution before the provided time limit.}
\label{table:heuristic2}
\end{table*}

\end{document}